\documentclass[prb,twocolumn,preprintnumbers]{revtex4}
\usepackage{graphicx}

\begin{document}

\title{ Quantum radiations from exciton condensate in Electron-Hole Bilayer Systems}
\author{Jinwu Ye$^{1}$, T. Shi$^{2}$, Longhua Jiang $^{1}$ and  C. P. Sun$^{2}$ }
\affiliation{$^{1}$Department of Physics, The Pennsylvania State
University, University Park, PA, 16802, USA   \\
 $^{2}$ Institute of Theoretical Physics, Chinese Academy of Sciences, Beijing,
100080, China }
\date{\today }

\begin{abstract}
Superfluid has been realized in Helium-4, Helium-3 and ultra-cold
atoms. It has been widely used in making high-precision devices and
also in cooling various systems. There have been extensive
experimental search for possible exciton superfluid (ESF) in
semiconductor electron-hole bilayer (EHBL)  systems below liquid
Helium temperature. Exciton superfluid are meta-stable and will
eventually decay through emitting photons. Here we  find that the
light emitted from the excitonic superfluid has unique and unusual
features not shared by any other atomic or condensed matter systems.
We show that  the emitted photons along the direction perpendicular
to the layer are in a coherent state with a single energy, those
along all tilted directions are in a two modes squeezed state. We
determine the two mode squeezing spectra, the angle resolved photon
spectrum, the line shapes of both the momentum distribution curve
(MDC) and the energy distribution curve (EDC). By studing the two
photon correlation functions, we find there are photon bunching, the
photo-count statistics is super-Poissonian. We also stress the
important difference between the quasi-particle excitations in an
equilibrium superfluid and those in a stationary state superfluid.
This difference leads to the explanation  of recent experimental
observation  of excitation spectrum of exciton-polariton inside a
planar cavity. We discuss how several important parameters such as
the chemical potential, the exciton decay rate, the quasiparticle
energy spectrum and the dipole-dipole interaction strength between
the excitons in our theory can be extracted from the experimental
data and comment on available experimental data on both EDC and MDC.
We suggest that all the predictions achieved in this paper can be
measured by possible future angle resolved power spectrum, phase
sensitive homodyne measurements, and HanburyBrown-Twiss type of
experiments. We demonstrate explicitly that the photoluminescence
from the exciton in EHBL systems  is a very natural, feasible and
unambiguous internal probe of the nature of quantum phases of
excitons in EHBL such as the ground state and the quasi-particle
excitations above the ground state. These remarkable features of the
photoluminescence can be used for high precision measurements,
quantum communication, quantum information processing and also for
the development of a new generation of powerful opto-electronic
devices.
\end{abstract}

\maketitle

\section{Introduction}

An Exciton is a bound state of  an electron and a hole. Exciton
condensate was first proposed more than 3 decades ago as a possible
ordered state in solids \cite{blatt}.  Keldysh and Kozlov \cite{kel}
argued that in a bulk semiconductor, in the dilute limit $n_{ex}
a^{3}_{ex} \ll 1 $ where $n_{ex} $ is the exciton density and
$a_{ex} $ is the exciton radius, the excitons behave as weakly
interacting bosons, the exciton effective mass $ M $ is even smaller
than an electron mass, for experimentally accessible exciton
densities, the 3 dimensional Bose-Einstein condensation (BEC)
critical temperature can be estimated to be $\sim K $. So in
principle, the excitons can undergo BEC and
become an excitonic superfluid state below a few $K $. In the dense limit $%
n_{ex} a^{d}_{ex} \gg 1 $, the fermionic nature of the electrons and holes
in the exciton will show up, the strong pairing BEC superfluid will
crossover to weak pairing BCS superfluid \cite{old}. However, in reality, it
is very difficult to realize the BEC of excitons experimentally in a bulk
system, because the short lifetime $\tau_{ex} $ and the long lattice
relaxation time $\tau_{L} $ which is needed for the hot exciton gas to reach
the cold temperature of the underlying lattice by emitting longitudinal
acoustic phonons. Although exciton gas, bi-excitons and electron-hole plasma
(EHP) phase have been observed in different bulk semi-conductors, no exciton
superfluid phase has been observed in any bulk semi-conductors.

Recently, degenerate exciton systems have been produced by different
experimental groups with two different methods in
quasi-two-dimensional
semiconductor $GaAs/AlGaAs $ coupled quantum wells structure \cite%
{butov,snoke,field1,field2,bell}. When the distance between the two
quantum wells is sufficiently small, an electron in one well and a
hole in the other well could pair to form an exciton which behaves
as a boson in dilute exciton density limit ( Fig.1a). This kind of
inter-layer excitons are called in-direct excitons.
In Butov and Snoke's labs ( also Bell lab ) \cite{butov,snoke,bell},
the excitons are created by optical pumping and then a electric
field is applied along the $\hat{z} $ direction to separate
electrons from holes by a distance $d \sim 30 nm $.  There are also
the current efforts from Bell lab \cite{bell} which focused on the
effects of electrostatic traps to confine the excitons in a given
regime. In the undoped electron-hole bilayer (EHBL) sample prepared
in Mike Lilly' lab
\cite{field1,field2,excitonprl,imbgate,layerinter,mag} and the
Cambridge group \cite{camb1,camb2} which is a heterostructures
insulated-gate field effect transistors, separate gates can be
connected to electron layer and hole layer, so the densities of
electron and holes can be tuned independently by varying the gate
voltages. Low densities and high mobilities regimes for both
electrons and holes can be reached. Transport properties such as
Coulomb drag can be performed in this experimental set-up.

The quantum degeneracy temperature of a two dimensional excitonic superfluid
(ESF) can be estimated to be $T^{ex}_{d} \sim 3 K $ for exciton density $n
\sim 10^{10} cm^{-2} $ and effective exciton mass $m=0.22 m_{0} $ where $%
m_{0} $ is the bare mass of an electron, so it can be reached easily
by $He $ refrigerator. It was established that the indirect excitons
in EHBL has at least the following advantages over the excitons in
the bulk: (1) Due to the space separation of electrons and holes,
the lifetime $\tau_{ex} $ of the excitons is made to be $10^{3}\sim
10^{5} $ longer than that of those in bulk semi-conductors, now it
can be made as long as microseconds. Due to the relaxation of the
momentum conservation along the $\hat{z} $ direction, the
thermal lattice relaxation time $\tau_{L} $ of the indirect excitons can be made as $%
10^{-3} $ that of bulk excitons, now it can be made as short as
nanoseconds. So $\tau_{ex} \gg \tau_{L} $ is well satisfied even for
the direct semiconductor such as $GaAs $. (2) Because all the
electric dipoles are aligned normal to the 2d plane, the repulsive
dipole-dipole interaction is crucial to stabilize the excitonic
superfluid against the competing phases such as bi-exciton formation
and electron-hole plasma (EHP) phase. So EHBL is a very promising
system to observe BEC of in-direct excitons. There are two important
dimensionless parameters in the EHBL. One is the dimensionless
distance $\gamma= d/a_{B} $ ( $a_{B} $ is the Bohr radius ) between
the two layers. Another is $r_{s} $ where $r_{s} a_{B} $ is the
typical interparticle distance in a single layer. Recently, one of
the authors proposed that the EHBL maybe a more favorable system to
observe a metastable excitonic supersolid (ESS) than the Helium 4
system \cite{ye}. The global phase diagram at $T=0 $ labeled by the
two parameters is shown in Fig.3.

\begin{figure}
\includegraphics[width=7cm]{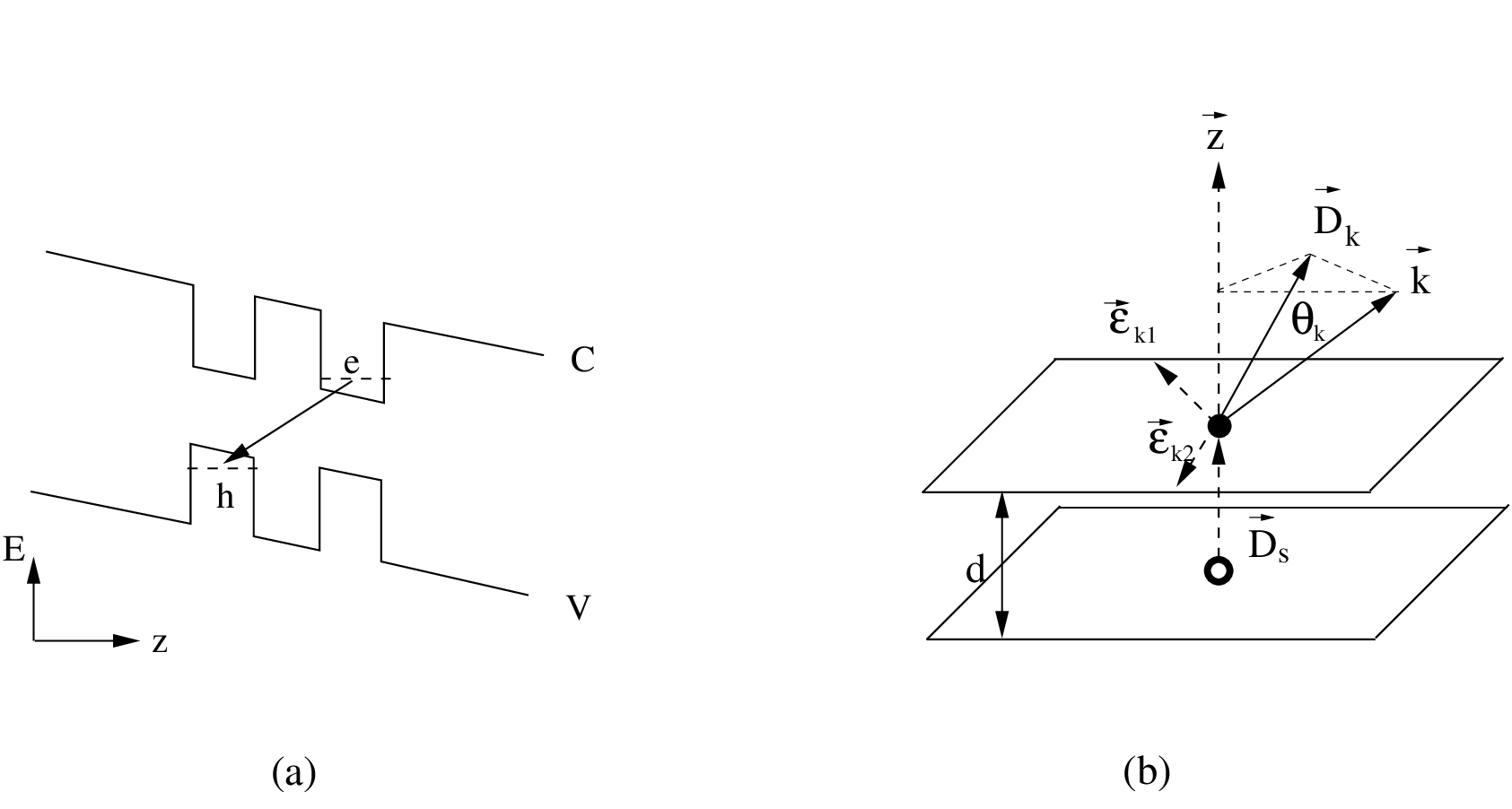}
\caption{ (a) In the experiments in \cite{butov,snoke,bell}, A laser
with excitation power $ P_{ex} $ is used to excite electrons from
the valence band ( V ) to the conduction band (C) in the
semiconductor GaAs/AlGaAs electron-hole bilayer systems. Then a gate
voltage $ V_{g} $ is applied along the $ z $ direction to separate
the electrons in the conduction band from the holes in the valence
band. One electron in one quantum well and one hole in the other
quantum well are bound to form an indirect exciton. In the
experiments in \cite{field1,field2}, the excitons are generated by
gate voltages. (b) The geometry of the photon emission from the
indirect excitons } \label{fig1}
\end{figure}

Note that in all the previous experiments \cite{butov,snoke,bell}, a
laser beam was consistently shined on a given 'bright' ring ( or a
'bright' spot ), however, the excitons will move to different
locations which is  at the center of the ring ( or a ring ) before
they annihilate and emit lights ( Fig.2 ). So in the stationary
process of emitting lights, the number of the exciton condensation
$N $ is kept to be a constant. The laser beam was used to
photo-generated the excitons, so it plays the role of a pump,
however, because the exciton condensation happens at different
locations than the laser pumping point, so if the BEC of excitons
indeed happens near the center of the trap, it is  indeed
spontaneous instead of being stimulated.  Indeed, as temperature
\cite{butov} is decreased from $ \sim 20 K $ to $ \sim 1.7 K $,  the
spatially and spectrally resolved PL peak density centering around
the gap in Fig.1a $ E_{g} \sim 1.545 eV $ increases, the exciton
cloud size decreases to $ L\sim 30 \mu m $, the peak width shrinking
to $ \sim 1 meV $ at the lowest temperature $ \sim 1.7 K $. All
these facts indicate the possible formation of exciton condensate
around $ 1.7K $.  In our theoretical analysis in this paper, we
assume the exciton cloud already reached the lattice temperature by
interacting with lattice acoustic phonons  within the thermal
lattice relaxation time $ \tau_{L} $ during its relaxation process
to the bottom of the trap, it also become a superfluid through
mutual dipole-dipole repulsive interaction and start to radiate
photons at the exciton lifetime $ \tau_{ex} \gg \tau_{L} $, then we
will calculate all the characteristics of the photons emitted from
the exciton superfluid. We will compare our theoretical results with
the experimental data in Sec.IX. The transient photoluminescence
from the excitons created by a short laser pulse will be discussed
in a separate publication.


\begin{figure}
\includegraphics[width=4cm]{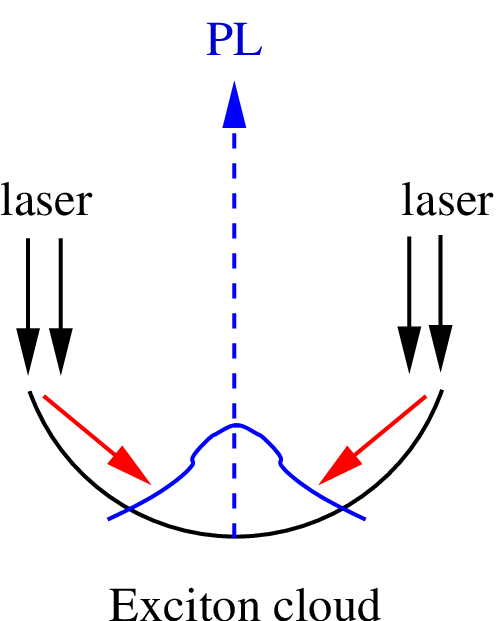}
\caption{ A laser ( black arrow ) with excitation power $ P_{ex} $
is used to excite electrons which are relaxed ( red arrow ) to the
bottom of the trap, then emit photons ( blue dashed arrow )
spontaneously.  } \label{fig2}
\end{figure}

    In parallel to search for exciton superfluid in EHBL,  extensive
    activities \cite{gold,hall,counterflow,blqhrev,counterflownature,fer,psdw,imb,cbtwo} have also been lavished on searching for exciton
    superfluid in electron-electron bilayer system  in the same
    semi-conductor material $AlGaAs/GaAs $ subject to a  high magnetic
    field in quantum Hall regime at total filling factor
    $ \nu_{T} =1 $.  When the interlayer separation $ d $ is sufficiently large, the bilayer
    system decouples into two separate compressible $ \nu=1/2 $ layers.
    However, when $ d $ is smaller than a critical distance $ d_{c} $,
    the system may undergo a quantum phase transition into
    a novel spontaneous interlayer coherent exciton superfluid phase \cite{blqhrev}.
    The exciton in this system can be considered as the
    pairing of an electron in top layer and the hole in the bottom
    layer after making a particle-hole transformation in the bottom
    layer. Other phases such as pseudo-spin density
    wave phase in some intermediate distance regimes was also proposed \cite{psdw,imb,cbtwo}.
    At low temperature, with extremely small interlayer tunneling amplitude,
  Spielman {\sl et al} discovered
  a very pronounced narrow zero bias peak in this possible exciton superfluid
  state \cite{gold}. M. Kellogg {\sl et al } also observed
  quantized Hall drag resistance at $ h/e^{2} $ \cite{hall}.
  In the counterflow experiments, it was found that
  both linear longitudinal and Hall resistances take activated
  forms and vanish only in the zero temperature limit \cite{counterflow}.
  However, despite the intensive theoretical research
  \cite{blqhrev} in the past,
  there are still many serious discrepancies between theory and the experiments.
  It remains unclear if the excitonic superfluid was indeed realized in the BLQH system.

 It is instructive to compare the measurements to detect possible exciton superfluid in the BLQH and EHBL
 In the BLQH, there are mainly three kinds of transport
experimental measurements (1) Quantum Hall resistance (2) Interlayer
tunneling (3) Counterflow.   In contrast to these quantum phases in
BLQH which are stable ones, all these excitonic phases in EHBL in
the Fig.3 are just meta-stable states which will eventually decay by
emitting lights. So the most natural experimental measurement for
photo generated EHBL is the photoluminescence (PL) which is quite
different from all the transport measurements in the BLQH. The
geometry of the photoluminescence from EHBL systems is shown in
Fig.1b. In fact, the photon emission in EHBL  plays a similar role
as the interlayer tunneling the BLQH, so the theoretical results
achieved in  both systems should shed on and transfer lights to each
    other. Very recently, transport experiments such as Coulomb drag
    were also performed in EHBL generated by gate
    voltages \cite{field1,field2,excitonprl,imbgate,layerinter,mag}.
    It is possible to also perform  counterflow experiment
    in the near future. The PL experiment can also be
    performed in this gate voltage generated EHBL, although the emitted lights are
    weaker than those from the photo generated EHBL \cite{private}.


\begin{figure}
\includegraphics[width=7cm]{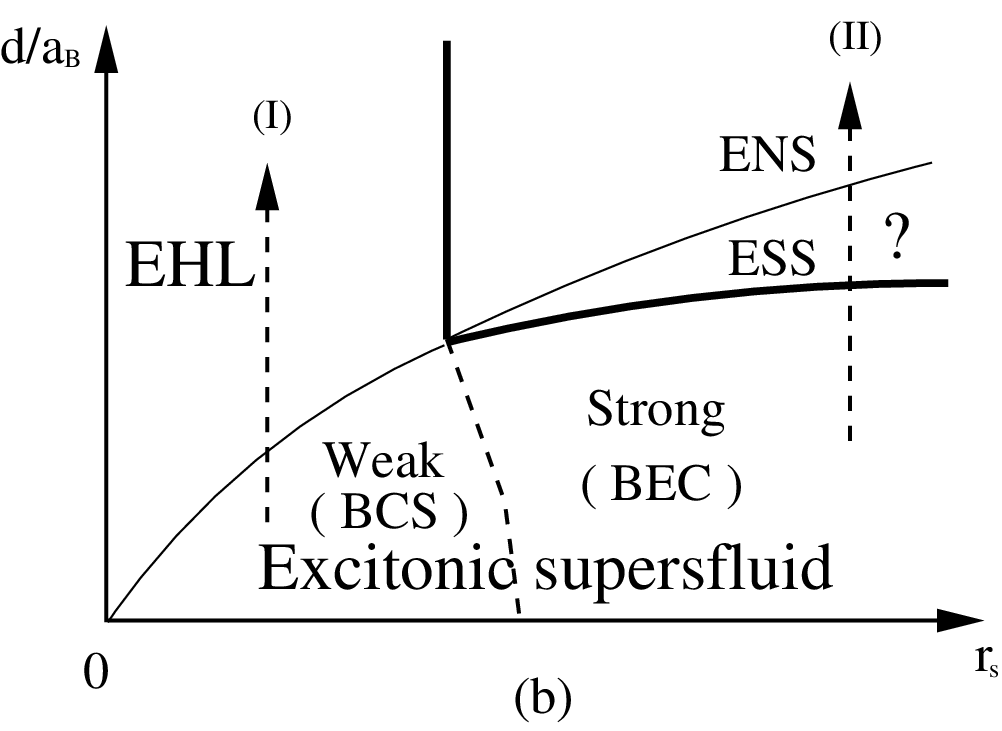}
\caption{ Zero temperature phase diagram of electron and hole
bilayer at the balanced case $n_{e}=n_{h}$ ( but $ m_e \neq m_h $ ).
The very important effects of spins, traps and disorders are not
taken into account. The $d$ is interlayer distance. $r_{s}a_{B}$
where $a_{B}$ is the Bohr radius is the average inter-particle
distance in a single layer. At high densities ( low $r_{s}$ ), along
the dashed line (I), there is a transition from the weak pairing (
or BCS ) ESF to the electron-hole plasma ( EHP ) phase. At low
densities ( high $r_{s}$ ), along the dashed line (II), there is a
transition from the strong pairing ( or BEC ) ESF to the excitonic
supersolid (ESS), then to excitonic normal solid ( ENS ). Inside the
ESF, there is a BCS ( weak-pairing ) to BEC ( strong pairing )
cross-over tuned by $r_{s}$ denoted by the dashed line. The EHP
phase is a conducting phase, while the ENS is an insulating phase,
so the EHP to ENS transition can be considered as a metal to
insulator transition. Thick line is 1st order transition, the thin
one is second order. The EHP to the weak pairing BCS transition may
also be 1st order. } \label{fig3}
\end{figure}

   In parallel to the experimental search for the exciton
    superfluid in the EHBL, there are also extensive experimental
    activities to search for exciton-polariton superfluid inside a
    planar micro-cavity. Although exciton condensation in a single quantum well ( SQW )
    has not been observed so far, there are
    some evidences for the observation of Exciton-polariton (EP) condensation in SWQ
    enclosed  inside a planar microcavity \cite{exp1,exp2,exp3,exppi,expp,expv,huig2,gan,gan2,laseing,revmicro}.
  These evidences include macroscopic occupation of the ground
  state, spectral and spatial narrowing, a peak at zero momentum in
  the momentum distribution ( see Fig.8b ) and spontaneous linear polarization of
  the light emission and so on. The elementary excitation spectrum
  of exciton-polariton was also found \cite{expp} to be very similar to that in
  an equilibrium superfluid with notable exceptions near $ k=0 $.
  This puzzle will be resolved in Sec.VII-6.
  Recently, several ultra-cold atom experiments
  \cite{qedbec01,qedbec02,qedbec1,qedbec2}
   successfully achieved the strong coupling of a BEC of $ N \sim 10^{5} $ $ ^{87}Rb $ atoms
   inside a cavity.
   Motivated by these achievements of SQW and atomic BEC embedded inside a cavity,
  we suggest that in near future experiments,  the EHBL can also be enclosed in
  a planar micro-cavity, so one can search for possible superfluid of indirect exciton polartion (IEP).
  One advantage of the EHBL over the SQW is that as shown in
  the previous paragraph, the dipoles of the indirect excitons are all aligned along the $
  \hat{z} $ direction, so the dipole-dipole interaction is
  repulsive, this also guarantees the IEP-IEP interaction is
  repulsive which is a sufficient and necessary condition to
  stabilize a superfluid against other possible states.
  This strong coupling regime in a planar micro-cavity
  will be investigated in separate publications.

Although there exist extensive experiment measurements on
photoluminescence from presumably achieved exciton BEC in the
electron-hole bilayer (EHBL) system \cite{field1,field2,bell}, so
far, there is no systematic theory on how photons interact with the
indirect excitons in different quantum phases in the EHBL system and
how the characteristics of photons can reflect the nature of the
quantum phases in the Fig.3. In this paper, we will study three
dimensional photons interacting with the two dimension indirect
excitons in the excitonic superfluid phase in the BEC side in the
EHBL.
 We will work out how the photoluminescence from this phase can reflect
the properties of both the condensate and the Bogoliubov
quasi-particle excitations above the condensate at zero temperature
$T=0 $. We find that due to the non-vanishing order parameter in the
ESF phase, the emitted photons along the direction perpendicular to
the layers ( namely with zero in-plane momentum ) are in a coherent
state, while the non-vanishing anomalous Green function in the ESF
lead to a two mode squeezed state of the emitted photon along all
tilted directions ( namely, at finite in-plane momenta ) as shown
Fig. 1b. We determine the angle resolved power spectrum, squeezing
spectrum, one and two photon correlation functions along all the
possible directions including normal and tilted directions. From the
two point correlation function, we can identify the quantum nature
of the emitted photons such as photon bunching, anti-bunching, also
the photo-count statistics such as super-Poissonian, Poissonian and
sub-Poissonian \cite{book1,book2}.
 We will also determine the momentum distribution curve (MDC) and
energy distribution curve (EDC) which are the integrated angle
resolved power spectrum at fixed energy and fixed momentum
respectively and then compare with the available experimental data
on EDC. We will suggest that all our predictions can be measured by
possible future angle resolved power spectrum, phase sensitive
homodyne measurements, and HanburyBrown-Twiss type of experiments.
We will also elucidate the physical reasons why the angle resolved
power spectrum takes the super-radiant form even in the
thermodynamic limit when the exciton decay rate is sufficiently
large, why the characters of the light emitted from the ESF phase
can reflect both the nature of the ground state and the Bogoliubov
quasi-particle excitations above the ground state even at $T=0 $.
The photoluminescence from the other phases in EHBL system will be
studied in subsequent works. In this paper, we did not consider the
very important effects of spins of electrons and holes which lead to
the formation of the bright excitons with $ J=\pm 1 $ and the dark
excitons with $ J=\pm 3/2 $ \cite{spin}, the effects of the trap,
finite thickness and  disorders in the sample \cite{trap}. All these
will be discussed in separate publications.

The rest of the paper was organized as follows, in section II, we
will derive the interaction between the 3 dimensional photons with 2
dimensional indirect excitons in the exciton gas phase in EHBL with
a distance $d $. In section III, we will derive the total
Hamiltonian in the ESF phase on the BEC side, separating the
interaction into the coupling to the condensate at zero in-plane
momentum $\vec{k}=0 $ and to the Bogolubov quasi-particle at
non-zero in-plane momentum $\vec{k}\neq 0 $. Then we will show how a
coherent state is emitted at $\vec{k}=0 $ and discuss several
remarkable properties of power spectrum along the normal direction
in section IV. In Section V, we will develop systematically an
input- output formalism for a stationary state. Then by using the
input-output formalism, we will calculate various emitted photon
characteristics in the follow sections. In Section VI, by
calculating the squeezing spectrum, we show that the ESF phase of
the excitons play a similar role as a two mode squeezing operator
which squeeze the input vacuum state into a two mode squeezed state,
so the emitted photons at non-zero $\vec{k} $ are always in a two
mode squeezed state even off the resonance. In section VII, we
evaluate the angle resolved power spectrum, the line shapes of both
MDC and EDC. We find that the angle resolved power spectrum takes a
stationary super-radiant form even in the thermodynamic limit when
the exciton decay rate is sufficiently large compared to the energy
of the Bogoliubov excitation. We also compare with the Dicke model
on super-radiance of finite $ N $ atoms in conventional quantum
optics. By working out the special nature of excitation spectrum in
an non-equilibrium superfluid, we resolve the puzzle observed in
\cite{expp}. We also resolve the In section VIII, we compute the one
and two photon correlation functions at non-zero in-plane momentum
$\vec{k}\neq 0 $ and find the photon statistics at any non-zero $
\vec{k}$. In section IX, we will compare our theoretical results on
EDC and MDC achieved in the last section with the previous
experimental PL data in \cite{butov,coherence} and also discuss
possible future experimental set-up such as angle resolved power
spectrum measurement, phase sensitive homodyne measurement, and
HanburyBrown-Twiss type of experiments to test the predictions
achieved in sections IV-VIII. In the final section X, we summarize
the main results on the coherent, squeezed and macroscopic
super-radiant nature of the emitted photons and point out their
crucial difference than the previous coherent and squeezed states
generated by pumps in non-linear media.  In the appendix A, we will
explain how the exciton superfluid emit photons in terms of a
intuitive Radiation Zone picture. In the appendix B which supplement
section VI, we give a more intuitive proof that all the emitted
photons along the tilted directions are in a two mode squeezed state
even off the resonance. In the appendix C, we clarify the relation
between the quantities calculated in the main text and experimental
measurable quantities. In the appendix D, we will perform a Golden
rule calculations to second order at both $ \vec{k}=0 $ and $
\vec{k} \neq 0 $ by using the many body exciton BEC ground and
excited states with Bogoliubov quasi-particles and compare with the
results achieved in section IV by Heisenberg equation of motion at $
\vec{k}=0 $ and in section VI-VIII by the non-perturbation
input-output formalism calculations at $ \vec{k} \neq 0 $.

\section{The interaction of exciton with photon in the BEC side of EHBL: a
microscopic point of view}

In this section, we will derive the coupling constant between the three
dimensional photons with two dimensional indirect excitons from microscopic
point of view. The second quantization Hamiltonian consists of three parts $%
H=H_{A}+H_{ex}+H_{int}$ where the first part is the Hamiltonian of free
photons:
\begin{equation}
H_{A}=\sum_{k\mathbf{,}\lambda }\omega _{k}a_{k,\lambda }^{\dagger
}a_{k,\lambda }  \label{first}
\end{equation}%
where $a_{k,\lambda }$ ($a_{k,\lambda }^{\dagger }$) is the
annihilation (creation) operator of the photon, it has polarization
$\lambda $ and three dimension momentum $k=(\vec{k},k_{z})$ where
$\vec{k}$ is the two dimensional in-plane momentum. The frequency of
the photon is $\omega _{k}=v_{g}\sqrt{k_{z}^{2}+\vec{k}^{2}}$, where
$v_{g}=c/\sqrt{\epsilon }$. Here, $c$ is the light speed in the
vacuum and $\epsilon \sim 12$ is dielectric constant of $GaAs$.

In EHBL, we can decompose the electron field into two parts:
\begin{equation}
\psi (\vec{r},z)=\psi _{1}(\vec{r},z)\Phi _{1}(z)+\psi _{2}(\vec{r},z)\Phi
_{2}(z).
\end{equation}%
where $ \vec{r} $ stands for two dimensional positions in the EHBL,
$\Phi_{1}(z) $ and $\Phi_{2}(z) $ are strongly localized around $z=0
$ and $z=d $ respectively. Then $\psi _{1}(\vec{r},z)\simeq \psi
_{1}(\vec{r},z=0) \equiv \psi _{1}(\vec{r})$ and $\psi
_{2}(\vec{r},z)\simeq \psi _{2}( \vec{r},z=d) \equiv \psi
_{2}(\vec{r})$ are electron operators in top and bottom layers in
Fig.1a.  The second part is the Hamiltonian of the exciton:
\begin{eqnarray}
H_{ex} &=&H_{0}+V_{int}  \nonumber \\
H_{0} &=&\int d^{2}r\psi _{1}^{\dagger }(\vec{r})[\frac{-\hbar
^{2}\vec{
\nabla}^{2}}{2m_{0}}+V_{1c}(\vec{r})]\psi _{1}(\vec{r})  \nonumber \\
&&+\int d^{2}r\psi _{2}^{\dagger }(\vec{r})[\frac{-\hbar
^{2}\vec{\nabla}^{2}
}{2m_{0}}+V_{2c}(\vec{r})]\psi _{2}(\vec{r})  \nonumber \\
V_{int} &=&\frac{1}{2}\int d^{2}rd^{2}r^{\prime }\delta \rho _{i}(\vec{r}%
)V_{ij}(\vec{r}-\vec{r}^{\prime })\delta \rho _{j}(\vec{r^{\prime
}}) \label{ehbl}
\end{eqnarray}%
where $V_{1c}$ and $V_{2c}$ are periodic crystal potentials in the
two layers, $\delta \rho _{i}(\vec{r})=\psi _{i}^{\dagger
}(\vec{r})\psi _{i}( \vec{r})-n_{i},i=1,2$ are normal ordered
electron densities on each layer. The intralayer interactions are
$V_{11}=V_{22}=e^{2}/\epsilon |\vec{r}| $, while interlayer
interaction is $V_{12}=V_{21}=e^{2}/\epsilon \sqrt{\left\vert
\vec{r}\right\vert ^{2}+d^{2}}$ where $\epsilon $ is the dielectric
constant.

Considering the effects of the crystal potentials $V_{1c}$ and $V_{2c}$ in
the two layers, the electron field operators in the two layers can be
expanded in terms of Bloch waves:%
\begin{eqnarray}
\psi _{1}(\vec{r}) &=&\frac{1}{\sqrt{S}}\sum_{\vec{k}}u_{c,\vec{k}}(\vec{r}%
,z_{1}=0)e^{i\vec{k}\mathbf{\cdot }\vec{r}}c_{1\vec{k}},  \nonumber \\
\psi _{2}(\vec{r}) &=&\frac{1}{\sqrt{S}}\sum_{\vec{k}}u_{v,\vec{k}}(\vec{r}%
,z_{2}=d)e^{i\vec{k}\mathbf{\cdot }\vec{r}}c_{2\vec{k}},
\end{eqnarray}%
where $S$ is the area of the layers and the Bloch wave functions satisfy $%
H_{0}u_{i,\vec{k}}(\vec{r})e^{i\vec{k}\mathbf{\cdot }\vec{r}}=E_{i}u_{i,\vec{%
k}}(\vec{r})e^{i\vec{k}\mathbf{\cdot }\vec{r}}$ with $ i=c,v $. For
direct semi-conductor
such as $GaAs$, there is a minimum $\epsilon _{i}$ in the conduction band $%
i=1$ and a maximum $\epsilon _{v}$ at the valance band $i=2$. We will set $%
\epsilon _{v}=0$ below, then the band gap is $E_{g}=\epsilon _{i}-\epsilon
_{v}=\epsilon _{v}$.

It is convenient to perform the particle-hole transformation in the valence
band $c_{1\vec{k}}=e_{\vec{k}}$ and $c_{2\vec{k}}=h_{-\vec{k}}^{\dagger }$,
then $e_{\vec{k}}$ ($h_{\vec{k}}$) is annihilation operator of the electron
(hole), then the exciton Hamiltonian can be rewritten as
\begin{equation}
H_{ex}=\sum_{\vec{k}}[E_{c}(\vec{k})e_{\vec{k}}^{\dagger }e_{\vec{k}}+E_{v}(%
\vec{k})h_{\vec{k}}^{\dagger }h_{\vec{k}}]+V_{ij},
\end{equation}

In the rest of the paper, we consider the dilute limit along the
path II in Fig.3 where the size of the exciton is of the order of
distance between the two layers. The interaction between electron
and electron (or hole and hole) in the same layer will just
renormalize the masses of the electron (hole) in the same layer
\cite{SS}. Then the Hamiltonian of the exciton can be further
simplified to%
\begin{equation}
H_{ex}=\sum_{\vec{k}}[(\frac{\vec{k}^{2}}{2m_{e}}+E_{g})e_{\vec{k}}^{\dagger
}e_{\vec{k}}+\frac{\vec{k}^{2}}{2m_{h}}h_{\vec{k}}^{\dagger }h_{\vec{k}%
}]+V_{e-h},
\end{equation}%
where $m_{e}$ ($m_{h}$) is the effective mass of the electron (hole). The
exciton creation operator is defined as $b_{\vec{k}}^{\dagger }=\sum_{\vec{p}%
}\varphi _{0}(\vec{p}-m_{e}\vec{k}\mathbf{/}M)e_{\vec{p}}^{\dagger }h_{\vec{k%
}-\vec{p}}^{\dagger }$, where the exciton mass $M=m_{e}+m_{h}$ and $\varphi
_{0}(\vec{p})$ is the Fourier transformation of the wave function $\phi _{0}(%
\vec{r})$ which satisfies%
\begin{equation}
\lbrack -\frac{1}{2m_{r}}\nabla _{\vec{r}}^{2}-\frac{e^{2}}{\epsilon \sqrt{%
\left\vert \vec{r}\right\vert ^{2}+d^{2}}}]\phi _{0}(\vec{r})=-E_{b}\phi
_{0}(\vec{r})  \label{bind}
\end{equation}%
where the reduced mass is $ 1/m_{r}=1/m_{e}+1/m_{h}$.

For a direct exciton in a single quantum well, $ d=0 $ in
Eqn.\ref{bind}, the size of an exciton is $a_{ex}=\hbar ^{2}
\epsilon /e^{2}m_{r} = \epsilon \frac{m_{0}}{m_{r}} a_{B} \sim 100
a_{B} \sim 50 \AA $ where $ a_{B}=\hbar ^{2} /e^{2}m_{0} \sim 0.53
\AA $ is the bare Bohr radius with the $ S $-wavefunction $
\phi_{0}(r)= (\frac{8}{\pi a^{2}_{ex}})^{1/2} e^{- 2r/a_{ex}} $. The
binding energy was known to be $ E_{b}= -e^{2}/2a_{ex}\epsilon = -
\frac{m_{r}}{m_{0}} \frac{1}{ \epsilon^{2}}  \frac{ e^{2}}{ 2 a_{B}}
\sim -10 meV $. For an indirect exciton in the EHBL, $ d > 0 $, the
exact form of the solution of Eqn.\ref{bind} is not known, but it is
not needed in the following discussions.  Taking the exciton density
$ n \sim 10^{10} cm^{-2} $, we can see that the average spacing
between excitons $ a \sim 100 nm \gg a_{ex} \sim 5 nm $, so the
sample is in the dilute limit.
The exciton operators satisfy the commutation relation $[b_{\vec{k}},b_{\vec{%
k}}^{\dagger }]=1$ approximately in the dilute limit along the path
II in Fig.3. This approximation is valid when the electron and hole
form a tight bound state, the pair breaking process into electron
and hole is at very high energy and can be neglected at low
temperature, the excitonic system is essentially a bosonic system.
Finally the Hamiltonian of the free exciton reads
\begin{equation}
H^{0}_{ex}=\sum_{\vec{k}}E_{\vec{k}}^{ex}b_{\vec{k}}^{\dagger
}b_{\vec{k}}, \label{second}
\end{equation}%
where $E_{\vec{k}}^{ex}= \hbar^{2} \vec{k}^{2}/2M + E_{g}-E_{b}$.
The third part is the interaction between excitons and photons which
can be separated into one photon and two photon parts:
\begin{eqnarray}
H_{int}^{(a)} &=&-e/m_{0}\int d^{3}r\psi ^{\dagger }(r)\vec{A}(r)\cdot \vec{p%
}\psi (r),  \nonumber \\
H_{int}^{(b)} &=&e^{2}/2m_{0}\int d^{3}r\psi ^{\dagger }(r)\vec{A}%
^{2}(r)\psi (r),
\end{eqnarray}%
where $ r=( \vec{r}, z ) $ stands for the {\em three dimensional}
position, $m_{0}$ is the bare mass of an electron, the vector
potential of the photon is:
\begin{equation}
\vec{A}(r)=\sum_{k,\lambda }\vec{\epsilon}_{k\lambda }\sqrt{%
1/2\epsilon \omega _{k}V}(a_{k,\lambda }e^{ik\mathbf{\cdot }r}+a_{k,\lambda
}^{\dagger }e^{-ik\mathbf{\cdot }r})
\end{equation}
where  $ V= L^{2} \times L_{z} $ is the normalization volume of the
whole 3-dimensional system and the $\vec{\epsilon}_{k\lambda }$ is
polarization of the photon with {\em three dimensional} momentum $k$
.

By inserting the vector potential $\vec{A}(r)$ and the electron (hole) field
into the interaction Hamiltonian $H_{int}^{(b)}$, the approximate relation
\cite{SS} $\int d^{2}\vec{r} u_{c,0}^{\ast }(\vec{r})u_{v,0}(\vec{r})\simeq 0$ leads to $%
H_{int}^{(b)}\simeq 0$ in the Hilbert space of the exciton. Then we
can project the interaction Hamiltonian $H_{int}^{(a)}$ into the
Hilbert space of the excitons $ \left\vert \vec{k}_{ex}\right\rangle
 =b_{\vec{k}}^{\dagger }\left\vert 0_{ex}\right\rangle $ where the $\vec{k}$
is the two dimensional momentum of the exciton. In this subspace of
the exciton, the interaction Hamiltonian is
\begin{eqnarray}
H_{int}^{(a)} &\simeq &\sum_{\vec{k}}[\left\langle \vec{k}_{ex}\right\vert
H_{int}^{(a)}\left\vert 0_{ex}\right\rangle \left\vert \vec{k}%
_{ex}\right\rangle \left\langle 0_{ex}\right\vert +h.c.]  \nonumber \\
&\simeq &\sum_{\vec{k}}[\left\langle \vec{k}_{ex}\right\vert
H_{int}^{(a)}\left\vert 0_{ex}\right\rangle b_{\vec{k}}^{\dagger
}+h.c.].
\end{eqnarray}%
By utilizing the electron field operator and the relation $[\vec{r}\mathbf{,}%
H_{0}]=i\vec{p}/m_{0}$, we find the matrix element $\left\langle \vec{k}%
_{ex}\right\vert H_{int}^{(a)}\left\vert 0_{ex}\right\rangle
=i\sum_{k_{z}}g(k)a_{k}$ where the coupling constant is
\begin{equation}
g(k)=-(1/2\epsilon \omega _{k}L_{z} )^{1/2}E_{\vec{k}}^{ex}\mu
_{cv}(k)\phi (0) \sim L^{-1/2}_{z} \label{lz}
\end{equation}%
where $ L_{z} $ is the normalization length along the $ z $
direction, $\mu _{cv}(k)=\vec{\epsilon}_{k \lambda }\cdot
\vec{D}_{k} $ where the transition dipole moment between the
conduction band and the valence band is
$\vec{D}_{k}=(\vec{D}_{xy},D_{z})$:
\begin{eqnarray}
\vec{D}^{k}_{xy } &=&\int d^{3}re^{ik_{z}z}u_{c,\vec{k}}^{\ast }(\vec{r},z)e\vec{r}%
u_{v,\vec{k}}(\vec{r},z) \Phi^{*}_{1}(z) \Phi _{2}(z),  \nonumber \\
D^{k}_{z} &=&\int d^{3}re^{ik_{z}z}u_{c,\vec{k}}^{\ast }(\vec{r},z)ezu_{v,\vec{k}}(\vec{r}%
,z) \Phi^{*}_{1}(z) \Phi _{2}(z).
\end{eqnarray}
 Both are essentially the overlap between the wavefunction of the
 electron in conduction band in one quantum well  and the
 wavefunction of the hole in valence band in the other quantum well which lead to a small
 interlayer tunneling.

For a given photon momentum with $  k $,  the polarization
$\vec{\epsilon}_{k2}= \vec{k} \times \vec{D}_k $ in Fig.1b is normal
to the transition dipole moment, so can be dropped out, we need only
consider the single polarization $ \vec{\epsilon}_{k1} $ which is in
the plane determined by $ \vec{k} $ and $ \vec{D}_k $ in Fig.1b,
then $\mu _{cv}(k)= D_{k} \sin \theta _{k}$. Note that the
transition dipole moment $ \vec{D}_{k} $ from the conduction band to
the valence band at a momentum $ k $ is completely different from
the static dipole moment $ \vec{D}_{s} = e \int d^{2} \vec{r} [
|\psi_{2} |^{2} - |\psi_{1} |^{2} ] $ in the dipole-dipole
interaction $V_{d}(\vec{q}) $ in Eqn. \ref{chem}. Although $
\vec{D}_{s} $  is completely along the z-direction in the dilute
limit along the path II in the Fig.3, the $\vec{D}_{k} $ is along a
general direction depending on $ k $ shown in Fig.1b. For example,
in the absence of interlayer tunneling, $ \vec{D}_{k} =0 $, but $
\vec{D}_{s} \neq 0 $.

Finally, the interaction Hamiltonian is simplified to%
\begin{equation}
H_{int}^{(a)}=\sum_{k}[ig(k)a_{k}b_{\vec{k}}^{\dagger }+h.c.],  \label{third}
\end{equation}
   where there is a in-plane momentum comservation between emitted
   photons and the excitons.

In the summary of this section, we derived the Hamiltonian $%
H=H_{A}+H_{ex}+H_{int}^{(a)}$ of the indirect exciton + photon Hamiltonian
in the EHBL system given by Eqns.\ref{first},\ref{second},\ref{third}.

\section{ The coupling between the photon and the condensate, the photon and the Bogoliubov quasi-particles
in the excitonic superfluid}

In this section, we will consider the effective interaction between
the photon and the Bogoliubov quasi-particle excitations in the
excitonic superfluid phase in the BEC side in Fig.4.
 The total Hamiltonian in
grand canonical ensemble is the sum of excitonic superfluid part,
photon part and the coupling between the two parts
$H_{t}=H_{sf}+H_{ph}+H_{int}$ where :
\begin{eqnarray}
H_{sf} &=&\sum_{\vec{k}}(E_{\vec{k}}^{ex}-\mu )b_{\vec{k}}^{\dagger }b_{\vec{%
k}}+\frac{1}{2 S } \sum_{\vec{k}\vec{p}\vec{q}}V_{d}(q)b_{\vec{k}-\vec{q}}^{\dagger }b_{%
\vec{p}+\vec{q}}^{\dagger }b_{\vec{p}}b_{\vec{k}}  \nonumber \\
H_{ph} &=&\sum_{k}\omega _{k}a_{k}^{\dagger }a_{k}  \nonumber \\
H_{int} &=&\sum_{k}[ig(k)a_{k}b_{\vec{k}}^{\dagger }+h.c.].
\label{chem}
\end{eqnarray}%
where $E_{\vec{k}}^{ex}= \hbar^{2} \vec{k}^{2}/2M + E_{g}-E_{b}$, $
S $ is the area of the sample,  $V_{d}(\vec{q})= \frac{ 2 \pi
e^{2}}{ \epsilon q } (1- e^{-qd} ) $ is the dipole-dipole
interaction between the excitons \cite{ye}, $V_{d}(\left\vert
\vec{r}\right\vert \gg d)=e^{2}d^{2}/\left\vert \vec{r}\right\vert
^{3}$ and $ V_{d}(0) = \frac{ 2 \pi e^{2} d}{ \epsilon } $ which is
a finite constant leading to a capacitive term for the density
fluctuation \cite{ye}. {\sl It is important to stress that in a
stationary state, the
    chemical potential $ \mu $ for the excitons in Eqn.\ref{chem} is
    kept fixed by the off-resonant pumping which is the laser
    pumping in \cite{butov,snoke,bell,light} and the gate voltage
    pumping in  \cite{field1,field2,excitonprl,imbgate,layerinter,mag}.
    Very similar point was also stressed in \cite{andrei} in the context of
    non-equilibrium stationary transport through a quantum dot.}

In the dilute limit, $V_{d}$ is relatively weak, so we can apply
standard Bogoloubov approximation to this system, this is in
contrast to Helium 4 system which is a strongly interacting system.
The phase $\theta $ representation used in
\cite{ye} is very useful to study vortex anti-vortex excitations and
Kosterlitz-Thouless transition at finite temperature. In this paper,
we focus only at $T=0$, so we can ignore the topological excitations
in the phase winding $\theta $ and just use the Bogoloubov
approximation to treat the non-topological low energy excitations.
So in the ESF phase, one can decompose the exciton operator into the
condensation part and the quantum fluctuation part above
the condensation $b_{\vec{k}}=\sqrt{N}\delta _{\vec{k}0}+\tilde{b}_{\vec{k}}$%
. Note that we ignored the zero point fluctuation above the
  condensate which is justified in the thermodynamic limit.
  In the real experimental situation where  finite number
  of excitons are trapped inside a trap, its importations will be
  addressed in a separate publication \cite{diffusion}.

  In the stationary state, the chemical potential is fixed at
\begin{equation}
\mu =E_{0}^{ex}+ \bar{n} V_{d}(0)= ( E_g-E_b) + \bar{n} V_{d}(0)
\label{fixed}
\end{equation}%
which is determined by eliminating the linear term of
$\tilde{b}_{0}$ in the Hamiltonian $H_{sf}$. So the chemical
potential is increased ( or called " blue shifted" ) from the single
exciton energy $ E_g-E_b $ by the dipole-dipole interaction $
\bar{n} V_{d}(0) $. In the experimental set-up shown in the Fig.1a,
$ E_g-E_b=E^{0}_g-E_b-eEd $ where $ E^{0}_g-E_b$ is the bare
conduction-valence band gap at zero gate voltage $ V_{g}=0 $ and $ E
$ is the electric field due to the applied voltage $ V_{g} $ in
Fig.1b. The exciton density $ \bar{n} $ is determined by the laser
excitation power $ P_{ex} $. So the chemical potential can be tuned
by the two experimental parameters $ V_{g} $ and $ P_{ex} $.

  Then the Hamiltonian of exciton BEC upto the
quadratic terms is
\begin{equation}
H_{sf}=\sum_{\vec{k}}[(\epsilon _{\vec{k}}+V_{d}(\vec{k})\bar{n})\tilde{b}_{%
\vec{k}}^{\dagger }\tilde{b}_{\vec{k}}+(\frac{V_{d}(\vec{k})\bar{n}}{2}%
\tilde{b}_{\vec{k}}^{\dagger }\tilde{b}_{-\vec{k}}^{\dagger }+h.c.)],
\end{equation}%
where the density of the condensate $\bar{n}=N/S$. For studying the
quasi-particle excitation spectrum of the exciton BEC, we utilize the
Bogoliubov transformation%
\begin{equation}
\beta _{\vec{k}}=u_{\vec{k}}\tilde{b}_{\vec{k}}+v_{\vec{k}}\tilde{b}_{-\vec{k%
}}^{\dagger } \label{uv}
\end{equation}%
to diagonize the Hamiltonian $H_{sf}$, where the transformation coefficients
are%
\begin{eqnarray}
u_{\vec{k}}^{2} &=&\frac{\epsilon _{\vec{k}}+\bar{n}V_{d}(\vec{k})}{2E(\vec{k%
})}+\frac{1}{2},  \nonumber \\
v_{\vec{k}}^{2} &=&\frac{\epsilon _{\vec{k}}+\bar{n}V_{d}(\vec{k})}{2E(\vec{k%
})}-\frac{1}{2} \label{factor}
\end{eqnarray}
 with $ u_{\vec{k}}^{2}-v_{\vec{k}}^{2}=1 $ and
 $ u_{\vec{k}} v_{\vec{k}}= \frac{ \bar{n} V_{d}(\vec{k}) }{ 2 E(\vec{k}) } $ which is completely due
 to the exciton dipole-dipole interaction. The quasi-particle creation and annihilation operators
  $\beta_{\vec{k}}$ and $\beta^{\dagger}_{\vec{k}^{\prime}} $ satisfy the Bose commutation relation $%
[\beta_{\vec{k}},\beta^{\dagger}_{\vec{k}^{\prime}}]=\delta_{\vec{k},
\vec{k}^{\prime} } $. Finally, the Hamiltonian of exciton BEC is
given by
\begin{equation}
H_{sf}=E(0)+\sum_{\vec{k}}E(\vec{k})\beta _{\vec{k}}^{\dagger }\beta _{\vec{k%
}}  \label{beta}
\end{equation}%
in terms of the quasi-particle creation and annihilation operators $\beta _{%
\vec{k}}$ and $\beta _{\vec{k}}^{\dagger }$ and $E(0)$ is the condensation
energy. The spectrum of quasi-particle excitation is%
\begin{equation}
E(\vec{k})=\sqrt{\epsilon _{\vec{k}}[\epsilon _{\vec{k}}+2\bar{n}V_{d}(\vec{k%
})]}, \label{qspectrum}
\end{equation}
As $\vec{k}\rightarrow 0$, $E(\vec{k})=u \left\vert \vec{k}%
\right\vert $ where the velocity of the quasi-particle is:
\begin{equation}
u=\sqrt{ \bar{n}V_{d}(0)/M} = \sqrt{ \frac{ 2 \pi e^{2} d \bar{n} }{
\epsilon M } }
 \label{sound}
\end{equation}

 Plugging $ \bar{n} \sim 10^{10} cm^{-2}, d \sim 30 nm, M \sim 0.22
 m_{0} $ into Eqn.\ref{sound}, we find $ u \sim 5 \times 10^{5} cm/s $.
 The quasi-particle spectrum is shown in Fig. 3 where the roton
mode is due to the long-range dipole-dipole interaction \cite{ye}.
Even at $ T=0 $, the number of excitons out of the condensate is:
\begin{equation}
 n^{\prime}(T=0) = \frac{1}{S} \sum_{\vec{k}}
 \langle \tilde{b}^{\dagger}_{\vec{k}} \tilde{b}_{\vec{k}} \rangle =
  \int \frac{d^{2} \vec{k} }{ (2 \pi)^{2} } v_{\vec{k}}^{2}
\label{dep}
\end{equation}
   which is the quantum depletion of the condensate due to the dipole-dipole
   interaction. From Eqn.\ref{factor},  we can see $ v_{\vec{k}}^{2}
   \rightarrow 1/k $, as $ k \rightarrow 0 $ and $ \rightarrow 1/k^{6} $, as $ k \rightarrow
   \infty $, so Eqn.\ref{dep} is well defined.

For studying the effects of the condensate and the quasi-particle excitation
to the emitted light separately, we decompose the interaction Hamiltonian $%
H_{I}$ into the coupling to the condensate part%
\begin{equation}
H_{I}^{c}=\sum_{k_{z}}[ig(k_{z})\sqrt{N}a_{k_{z}}+h.c.],  \label{cf1}
\end{equation}%
  and the coupling to the quasi-particle part:
\begin{equation}
H_{I}^{q}=\sum_{k}[ig(k)a_{k}\tilde{b}_{\vec{k}}^{\dagger }+h.c.].
\label{cf2}
\end{equation}%
In the following two sections, we discuss the properties of emitted
photons with the zero in-plane momentum $\vec{k}=0$ and the non-zero
in-plane momentum $\vec{k}\neq 0$ respectively. Note that due
 to the electron-hole asymmetry, we do not expect there is a
 up-down  $ Z_{2} $ symmetry. However, for the simplicity of
 notations, we assume there is such a $ Z_{2} $ symmetry, so we can
 treat the radiations in the upper and down half space on the equal
 footing. All our calculations can be generalized straightforwardly
 to take into account the asymmetry quantitatively in the real
 EHBL system.

{\sl In the following, in  order to keep the relative energy
difference between the exciton and the photon intact, we made a
rotation $a=  \tilde{a}(t)e^{-i \mu t} $, so we can focus on the
slowly varying $ \tilde{a}(t)$  and neglect the $ \tilde{} $ in the
following sections. }

\begin{figure}[tbp]
\includegraphics[bb=115 162 509 554, width=6cm, clip]{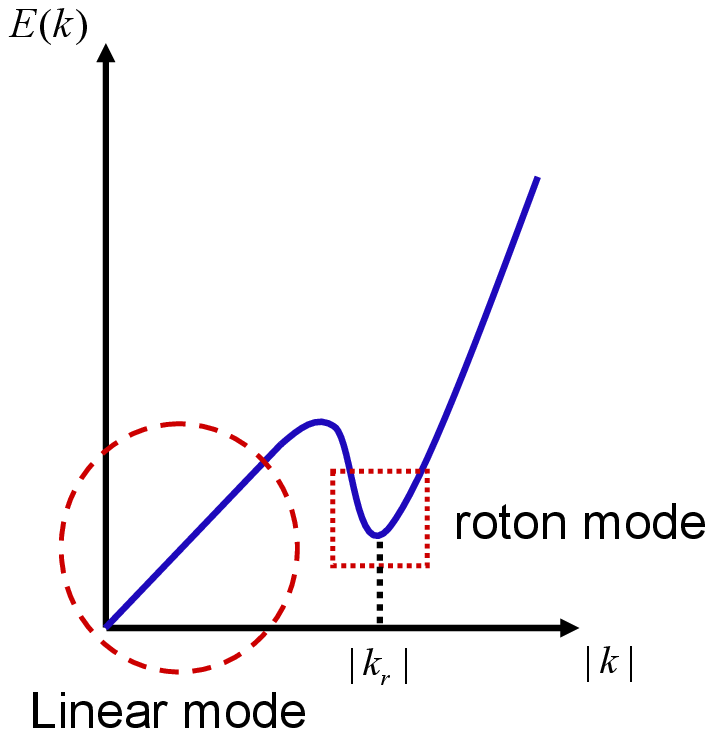}
\caption{The spectrum of the quasi-particle in exciton BEC where
there exist the linear mode and roton mode. This spectrum only holds
when teh lifetime of the exciton $ \tau_{ex} \rightarrow \infty $.
The spectrum  will be changed to Fig.6b when $ \tau_{ex} $ is large,
but finite.} \label{fig4}
\end{figure}
\section{The coherent state, line width and power spectrum in the normal direction at $ \vec{k}=0 $. }

Because the condensate carries no in-plane momentum $\vec{k}=0$, so from $%
H_{int}^{c}$ in Eqn.\ref{cf1}, the Heisenberg equation of motion of the
photon annihilation operator is%
\begin{equation}
i\partial _{t}a_{k_{z}}=(\omega _{k_{z}}-\mu -i\kappa %
)a_{k_{z}}-ig^{\ast }(k_{z})\sqrt{N}+iF(t)  \label{zero}
\end{equation}%
where we have dropped the zero mode fluctuation of the zero momentum
condensate which is negligible in the thermodynamic limit. The
$\kappa $ is the decay rate of the photon due to its coupling to a
reservoir ( or bath ), its value and physical meaning will be
determined self-consistently in the following. The  $F(t)$ is the
fluctuation of the reservoir satisfying $\left\langle b\right\vert
F(t)\left\vert b\right\rangle =0$ where $\left\langle O\right\rangle
_{b}=Tr(\rho _{b}O)$ denotes the mean value of the operator $O $ at
the bath state and $\rho _{b}$ is the density matrix of bath. From
Eqn.\ref{zero}, one can see the exciton condensation $\sqrt{N}$
plays the role of an effective pump on the photon part. As shown in
the Eqn.\ref{lz}, $ g(k_{z}) \sim L^{-1/2}_{z} $.

Following the standard laser theory, we decompose the operator as
its mean plus its fluctuation: $a_{k_{z}}=\left\langle
a_{k_{z}}\right\rangle _{in}+\delta a_{k_{z}}$ where the initial
state $\left\vert {}\right\rangle _{in}$ in Fig.2  is taken to be
$\left\vert BEC\right\rangle \left\vert 0\right\rangle_{ph}
\left\vert b\right\rangle $. Here, $\left\vert BEC\right\rangle $
denotes the ground state of the condensation; the initial photon
state $\left\vert 0\right\rangle_{ph} $ satisfies
$a_{k_{z}}(0)\left\vert 0\right\rangle_{ph} =0$ where $a_{k_{z}}(0)$
is the photon operator $a_{k_{z}} $ at the initial time $t=0$;
$\left\vert b\right\rangle $ denotes the state of the photon
reservoir. From Eqn.\ref{zero}, it is easy to see that
\begin{equation}
i\partial _{t}\left\langle a_{k_{z}}\right\rangle =(\omega _{k_{z}}-\mu -i%
\kappa )\left\langle a_{k_{z}}\right\rangle -ig^{\ast }(k_{z})%
\sqrt{N}.  \label{mean}
\end{equation}%
The stationary solution is%
\begin{equation}
\left\langle a_{k_{z}}\right\rangle =\frac{ig^{\ast }(k_{z})\sqrt{N}}{%
(\omega _{k_{z}}-\mu -i\kappa )} \label{ave}
\end{equation}
 which is the photon condensation  induced by the exciton
 condensation at $ \vec{k}=0 $. So the output state along the normal
 direction is a coherent state.

The fluctuation $\delta a_{k_{z}}$ obeys:
\begin{equation}
i\partial _{t}\delta a_{k_{z}}=(\omega _{k_{z}}-\mu -i\kappa )\delta
a_{k_{z}}+iF(t), \label{n1}
\end{equation}
When $t>>1/\kappa $, the photon fluctuation is completely determined by the
fluctuations of the reservoir:
\begin{equation}
\delta a_{k_{z}}(t)=\int_{0}^{t}d\tau F(\tau )e^{-i(\omega
_{k_{z}}-\mu )(t-\tau )-\kappa (t-\tau )}.  \label{n2}
\end{equation}
   Where the fluctuation-dissipation
relation dictates that $\left\langle F^{\dagger }(t)F(t^{\prime
})\right\rangle = \kappa \bar{n}_{\omega _{k_{z}}} \delta
(t-t^{\prime }), \left\langle F(t)F^{\dagger}(t^{\prime
})\right\rangle = \kappa ( \bar{n}_{\omega _{k_{z}}} + 1 ) \delta
(t-t^{\prime }), \left\langle F(t)F(t^{\prime })\right\rangle=
\left\langle F^{\dagger}(t)F^{\dagger}(t^{\prime })\right\rangle =0
$.

 From Eqn.\ref{n2}, we find:
\begin{equation}
\left\langle \delta a_{k_{z}}^{\dagger }(t)\delta
a_{k_{z}}(t^{\prime })\right\rangle = \frac{1}{2}
\bar{n}_{k_{z}}e^{- \kappa \left\vert t-t^{\prime }\right\vert }
e^{i(\omega_{k_z} -\mu )(t-t^{\prime })} \label{12}
\end{equation}

The power spectrum of the fluctuation is given by the Fourier
transformation of the correlation function:
\begin{equation}
S(\omega )=\int_{-\infty }^{+\infty }\left\langle \delta
a_{k_{z}}^{\dagger }(t)\delta a_{k_{z}}(t^{\prime })\right\rangle
e^{-i(\omega -\mu )(t-t^{\prime })}dt \label{n3}
\end{equation}
 where we have defined the
photon frequency with respect to the chemical potential $\mu $.

 By inserting Eqn. \ref{12} into  Eqn.\ref{n3}, we get and the power spectrum
\begin{equation}
S(\omega )= \frac{1}{2}  \frac{\bar{n}_{\omega _{k_{z}}}   \kappa
}{(\omega -\omega_{k_z} )^{2}+\kappa ^{2}},  \label{n4}
\end{equation}%
where the particle distribution of the photon reservoir is
$\bar{n}_{\omega }=1/(e^{\omega /T}-1)$ and $T$ is the temperature
of the reservoir. The result Eqn.\ref{n4} is consistent with the
Wiener-Khintchine theorem.

The number of the emitted photon is%
\begin{equation}
n_{\omega _{k_{z}}}=\left\langle a_{k_{z}}^{\dagger }a_{k_{z}}\right\rangle =%
\frac{N\left\vert g(\mu /c)\right\vert ^{2}}{(\omega _{k_{z}}-\mu
)^{2}+\kappa ^{2}}+\bar{n}_{\omega _{k_{z}}}/2, \label{nph}
\end{equation}%
where we have set $
g^{\ast }(k_{z})$ around $\omega _{k_{z}}=\mu $. Because the
condensate $N$ is much larger than the particle distribution of the
bath $\bar{n}_{\omega _{k_{z}}}$, so when the temperature of the
reservoir $ T \rightarrow 0 $ which is the case considered in this
paper, the number of the emitted photon is dominated by the first
term.

  Now we will determine the value of $ \kappa $ self-consistently.
  The total number of photons is
\begin{equation}
N_{ph}= \sum_{k_z  }n_{\omega_{k_{z}}}= N (|g|^{2}D )/\kappa
\label{npht0}
\end{equation}%
  where $ D= L_{z}/v_{g} $ is the
  photon density of states at $ \vec{k}=0 $.
  Note that the exciton decay rate $ \gamma_0 =|g|^{2}D $ at $ \vec{k}=0 $
  is independent of $ L_{z} $ ( see Eqn. \ref{dos} and \ref{decay} for general expressions of
  the density of state $ D_{\vec{k}}( \omega_{k} ) $
  and the exciton decay rate $ \gamma_{\vec{k}} $ at any $ \vec{k} $ ).
  The $ N_{ph} $ has to be proportional to $ L_{z} $ in order to get
  a finite photon density in a given volume $ L^{2} \times L_{z} $
  in a stationary state. This self-consistency condition sets $
  \kappa= v_{g}/L_{z} \rightarrow 0 $ so that
\begin{equation}
  N_{ph}=N \gamma_{0}L_{z}/v_{g} \sim L_{z}
\label{npht}
\end{equation}%
  Plugging this value of $ \kappa $ into Eqn.\ref{nph} leads to:
\begin{equation}
n_{\omega _{k_{z}}}= N \gamma_{0} \delta( \omega_{k_z}- \mu)
\label{nph1}
\end{equation}%
  which is independent of $ L_{z} $ as required !
  We showed that the power
  spectrum emitted from the exciton condensate has zero width.
  This conclusion is robust and is
  independent of any macroscopic details such as how photons are
  coupled to reservoirs.

  The radiation rate from the condensate is:
\begin{equation}
   P^{rd}_{0}= N \gamma_{0} \mu
\label{power0}
\end{equation}
   which is also independent of $ L_{z} $ as required.  In fact, in the limit $ \kappa=
   v_{g}/L_{z} \rightarrow 0 $, Eqns. \ref{n4}
   becomes $ S( \omega ) =  \bar{n}_{\omega_{k_z}} \delta ( \omega -
    \omega_{k_z} )$ which is negligible at low temperature. From the  Eqn.\ref{mean}, one can see that
   both the pumping term $ i g^{*} \sqrt{N} \sim L^{-1/2}_{z} $ and
   the dumping term $ i \kappa \sim L^{-1}_{z} $ approach zero as $
   L_{z} \rightarrow \infty $ limit in such a way that a stationary state
   is reached.

   From the experimental data in section IX, taking $ N\sim 10^{5},
   \gamma_{0} \sim 0.1 \mu eV, \mu \sim 1.54 eV $, we find
   $ P^{rd}_{0} \sim 1 \mu W $.

  In summary, the coherent light emitted from the condensate has the
  following remarkable properties: (1) highly directional: along the
  normal direction (1) highly monochromatic: pinned at a single
  energy given by the chemical potential $ \mu $ (3) high power:
  proportional to the total number of excitons.
  These remarkable properties are independent of any microscopic
  details as such as the excitation power $ P_{ex} $ and the line width of the pumping laser
  as long as they can generate excitons across the band gap.
  This fact could be useful to build highly
  powerful opto-electronic device. In the appendix D, we will
  give a more intuitive derivation of these results from a golden
  rule calculation.



\section{ The input-output formalism for a stationary state at $\vec{k}\neq 0$}

In this section, we consider the photons with in-plane momentum
$\vec{k}\neq 0$.   From Eqn.\ref{cf2}, it is easy to see that due to
the in-plane momentum conservation, the exciton with a fixed
in-plane momentum $\vec{k}$ coupled to 3 dimensional photons with
the same $\vec{k}$, but with different momenta $k_{z}$ along the
$z$-direction, so we can view these photon acting as the bath of the
exciton by defining $\Gamma _{\vec{k}}=\sum_{k_{z}}g(k)a_{k}$. As
shown in Eqn.\ref{dep},  due to the dipole-dipole repulsion, even at
$ T=0 $, there are also excitons depleted from the condensate. These
excitons will emit photons at non-zero $ \vec{k} $. By using the
standard input-output formalism for a stationary state discussed in
\cite{book1}, we will achieve the squeezed spectrum, angle resolved
power spectrum and photon correlation functions of the emitted
photon in the following sections.

The Heisenberg equations of motions of the photons and excitons are%
\begin{eqnarray}
\partial _{t}a_{k} &=&-i(\omega _{k}-\mu )a_{k}-g^{\ast }(k)\tilde{b}_{\vec{k%
}},  \nonumber \\
\partial _{t}B_{\vec{k}} &=&-i\Sigma B_{\vec{k}}+A_{\vec{k}},  \label{both}
\end{eqnarray}%
where $B_{\vec{k}}=(\tilde{b}_{\vec{k}},\tilde{b}_{-\vec{k}}^{\dagger })^{T}$%
, $A_{\vec{k}}=(\Gamma _{\vec{k}},\Gamma _{-\vec{k}}^{\dagger })^{T}$ and
\begin{equation}
\Sigma =\left(
\begin{array}{cc}
\epsilon _{\vec{k}}+\bar{n}V_{d}(\vec{k}) & \bar{n} V_{d}(\vec{k}) \\
-\bar{n} V_{d}(\vec{k}) & -\epsilon _{\vec{k}}-\bar{n}V_{d}(\vec{k})%
\end{array}%
\right) .
\end{equation}%
The formal solution of $a_{k}$ can be written either as the initial state at
$t_{0}<t$ or the final state at $t_{1}>t$:
\begin{eqnarray}
a_{k}(t) &=&a_{k}(t_{0})e^{-i(\omega _{k}-\mu )(t-t_{0})}  \nonumber \\
&&-g^{\ast }(k)\int_{t_{0}}^{t}dt^{\prime
}\tilde{b}_{\vec{k}}(t^{\prime })e^{-i(\omega _{k}-\mu )(t-t^{\prime
})} \label{phoi}
\end{eqnarray}%
and%
\begin{eqnarray}
a_{k}(t) &=&a_{k}(t_{1})e^{-i(\omega _{k}-\mu )(t-t_{1})}  \nonumber \\
&&+g^{\ast }(k)\int_{t}^{t_{1}}dt^{\prime
}\tilde{b}_{\vec{k}}(t^{\prime })e^{-i(\omega _{k}-\mu )(t-t^{\prime
})},  \label{phoo}
\end{eqnarray}

When plugging Eqns. \ref{phoi} and \ref{phoo} into Eqn.\ref{both},
we find it is convenient to define the input and output fields as:
\begin{eqnarray}
a_{\vec{k}}^{in}(t) &=&\sum_{k_{z}}\frac{1}{\sqrt{D_{\vec{k}}(\omega _{k})}}%
a_{k}(t_{0})e^{-i(\omega _{k}-\mu )(t-t_{0})},  \nonumber \\
a_{\vec{k}}^{out}(t) &=&-\sum_{k_{z}}\frac{1}{\sqrt{D_{\vec{k}}(\omega _{k})}%
}a_{k}(t_{1})e^{-i(\omega _{k}-\mu )(t-t_{1})},  \label{io}
\end{eqnarray}%
where the density of states of the photon with a given in-plane momentum $%
\vec{k}$ is%
\begin{equation}
D_{\vec{k}}(\omega _{k})=\frac{\omega _{k}L_{z} }{v_{g}\sqrt{\omega
_{k}^{2}-v_{g}^{2}\left\vert \vec{k}\right\vert ^{2}}}   \label{dos}
\end{equation}%
which is proportional to $ L_{z} $. It can be shown that if $ v_{g}
|\vec{k}| \ll \mu $, the input and output fields obey the Bose
commutation relations:
\begin{equation}
[a_{\vec{k}}^{in}(t),a_{ \vec{k}^{\prime} }^{in \dagger
}(t^{\prime})]=[a_{ \vec{k}}^{out}(t),a_{ \vec{k}^{\prime}
}^{out\dagger }(t^{\prime})]=\delta_{\vec{k},\vec{k}^{\prime}
}\delta (t-t^{\prime}) \label{comm}
\end{equation}

In term of the input field $\mathbf{a}_{\vec{k}}^{in}(t)=(a_{\vec{k}%
}^{in}(t),a_{-\vec{k}}^{in\dagger }(t))^{T}$, the exciton operator $B_{k}$
obeys%
\begin{equation}
\partial _{t}B_{\vec{k}}=(-i\Sigma -\frac{\gamma _{\vec{k}}}{2})B_{\vec{k}}+%
\sqrt{\gamma _{\vec{k}}}\mathbf{a}_{\vec{k}}^{in},  \label{in}
\end{equation}%

 In terms of the output field $\mathbf{a}_{\vec{k}}^{out}(t)=(a_{\vec{k}%
}^{out}(t),a_{-\vec{k}}^{out\dagger }(t))^{T}$, it obeys
\begin{equation}
\partial _{t}B_{\vec{k}}=(-i\Sigma +\frac{\gamma _{\vec{k}}}{2})B_{\vec{k}}-%
\sqrt{\gamma _{\vec{k}}}\mathbf{a}_{\vec{k}}^{out},  \label{o}
\end{equation}%
where  the effects of photon-exciton coupling are completely encoded
in the exciton decay rate  $\gamma _{\vec{k}}=D_{\vec{k}}(\mu
)\left\vert g_{\vec{k}}(\omega _{k}=\mu )\right\vert ^{2}$:%
\begin{equation}
\gamma _{\vec{k}}=\frac{E_{\vec{k}}^{ex2}\left\vert D_{k}
\right\vert ^{2}\sin ^{2}\theta _{k}\phi ^{2}(0)}{2\epsilon v_{g}
(\mu ^{2}-v_{g}^{2}\left\vert \vec{k}\right\vert ^{2})^{1/2} },
\label{decay}
\end{equation}%
where $ \theta _{k} $ is the angle between the transition dipole
moment $ \vec{D}_{k} $ and the 3 dimensional  photon vector $ k $
shown in Fig.1b. Note that $ \gamma _{\vec{k}} $ is independent $
L_{z} $, so is an experimentally measurable quantity as shown in
section IX-1. From the rotational invariance in the Fig.1b, we can
conclude that $\gamma _{\vec{k}} \sim const. + | \vec{k} |^{2} $ as
$ \vec{k} \rightarrow 0 $ as shown in Fig.6..

When comparing Eqn.\ref{in} and \ref{o} with Eqn.\ref{zero}, we can
see that input photon field $ \sqrt{\gamma_{\vec{k}}}
\mathbf{a}_{\vec{k}}^{in} $ ( or the output photon field $
\sqrt{\gamma_{\vec{k}}} \mathbf{a}_{\vec{k}}^{out} $ ) which are
summation of continuous spectral of photons at a given in-plane
momentum $ \vec{k} $, but with different $ k_z $ as shown in
Eqn.\ref{io} plays the role of the reservoir $ F(t) $ for the
excitons, while  the exciton decay rate $ \gamma_{\vec{k}} $ due to
the photon-exciton coupling plays the role of $ \kappa $. However,
there is no similar source ( or pumping ) term like $ -i g^{*}( k_z
) \sqrt{N} $.

When deriving Eqn.\ref{in} and \ref{o}, we have assumed the density
of state $D_{\vec{k}}(\omega _{k})$ at a given in-plane momentum
$\vec{k}$ varies slowly around the characteristic frequency $\omega
_{k}=\mu $. Indeed as shown later in Figs.7.8.10, the maximum of
squeezing spectrum and power spectrum is very narrowly peaked around
$ \omega_{k} =\mu $, so it is reasonable to set $ \omega_{k}=\mu  $
in $D_{\vec{k}}(\omega _{k})$. This approximation is essentially a
Markov approximation which is valid only when the in-plane momentum
$\left\vert \vec{k}\right\vert $ is much  smaller than $\mu /v_{g}$
in Eqn.\ref{decay}. In fact, as to be shown in section VI-1,
 the maximum in-plane momentum $\left\vert \vec{k}_{max} \right\vert =\mu
 /v_{g}  $.
This is also the same approximation for the commutation relations
Eqn.\ref{comm} hold. However, when the emitted photon is along the
plane, namely with $k_{z}=0$, the Markov approximation becomes
in-valid. So all our following calculations are valid as long as the
emitted photons are not too close to along the $xy $ plane.

The relation%
\begin{equation}
\mathbf{a}_{\vec{k}}^{in}+\mathbf{a}_{\vec{k}}^{out}=\sqrt{\gamma _{\vec{k}}}%
B_{\vec{k}},  \label{re}
\end{equation}%
is derived from Eqn.\ref{in} and Eqn.\ref{o}. The Fourier
transformations of Eq. (\ref{in}), Eq. (\ref{o}) and Eq. (\ref{re})
lead to input-output relation:
\begin{equation}
\mathbf{a}_{\vec{k}}^{out}(\omega )=W^{-1}(\omega )W^{\ast }(\omega
)\mathbf{a}_{\vec{k}}^{in}(\omega )
\end{equation}
 where $W(\omega )=-\gamma _{%
\vec{k}}I/2+i(\omega I-\Sigma )$ and $\mathbf{a}_{\vec{k}}^{out}(\omega )$
and $\mathbf{a}_{\vec{k}}^{in}(\omega )$ are the Fourier transformation of $%
\mathbf{a}_{\vec{k}}^{out}(t)$ and $\mathbf{a}_{\vec{k}}^{in}(t)$. Then the
component $a_{\vec{k}}^{out}(\omega )$ of the output field $\mathbf{a}_{\vec{%
k}}^{out}(\omega )$ is related to the input fields by:
\begin{eqnarray}
a_{\vec{k}}^{out}(\omega ) &=&[-1+\gamma _{\vec{k}}G_{n}(\vec{k},\omega +i%
\frac{\gamma _{k}}{2})]a_{\vec{k}}^{in}(\omega )  \nonumber \\
&&+\gamma _{\vec{k}}G_{a}(\vec{k},\omega +i\frac{\gamma _{k}}{2})a_{-\vec{k}%
}^{in\dagger }(-\omega ),  \label{bout}
\end{eqnarray}%
where the normal Green function $G_{n}(\vec{k},\omega )$ and the anomalous
Green function $G_{a}(\vec{k},\omega )$ are%
\begin{eqnarray}
G_{n}(\vec{k},\omega ) &=&i\frac{\omega +\epsilon _{\vec{k}}+\bar{n}V_{d}(%
\vec{k})}{\omega ^{2}-E^{2}(\vec{k})},  \nonumber \\
G_{a}(\vec{k},\omega ) &=&\frac{i\bar{n}V_{d}(\vec{k})}{\omega ^{2}-E^{2}(%
\vec{k})},
\end{eqnarray}%
which are determined by the properties of the quasi-particle in the
exciton BEC. In fact, they are just the retarded Green functions
after making the analytic continuation $ i \omega_{n} \rightarrow
\omega + i \delta $ in the corresponding imaginary time Green
functions. The exciton decay rate $ \gamma_{\vec{k}} $ in the two
Green functions in Eqn.\ref{bout} just stand for the fact that the
excitons are decaying into photons. Note that the Fourier
transformation of the Eq. (\ref{io}) leads to
\begin{equation}
\omega =\omega _{k}-\mu
\end{equation}

 In fact, Eqn.\ref{bout} can be viewed as a $ S $ matrix relating
the input photon field at $ t_0 \rightarrow - \infty $ to the output
photon field at $ t_1 \rightarrow \infty $.
 In the following sections,  we will
calculate the squeezed spectrum, angle resolved power spectrum and
photon correlation functions of the emitted photons respectively..

\section{ Two mode Squeezing spectrum with $ \vec{k} \neq 0 $ }

Eqn. \ref{bout} suggests that the output field is the two mode
squeezed state between $ \vec{k}$ and $ -\vec{k} $, so it is
convenient to define $A_{\vec{k},\pm
}^{out}(\omega )=[a_{\vec{k}}^{out}(\omega )\pm a_{-\vec{k}}^{out}(\omega )]/%
\sqrt{2}$ and $A_{\vec{k},\pm }^{in}(\omega )=[a_{\vec{k}}^{in}(\omega )\pm
a_{-\vec{k}}^{in}(\omega )]/\sqrt{2}$. Then:
\begin{eqnarray}
A_{\vec{k},\pm }^{out}(\omega ) &=&[-1+\gamma _{\vec{k}}G_{n}(\vec{k},\omega
+i\frac{\gamma _{k}}{2})]A_{\vec{k},\pm }^{in}(\omega )  \nonumber \\
&&\pm \gamma _{\vec{k}}G_{a}(\vec{k},\omega +i\frac{\gamma _{k}}{2})A_{\vec{k%
},\pm }^{in\dagger }(-\omega )  \label{Aout}
\end{eqnarray}

The position and momentum ( quadrature phase ) operators of the output field
is defined by%
\begin{eqnarray}
X_{\pm } &=&A_{\vec{k},\pm }^{out}(\omega )e^{i\phi _{\pm }(\omega )}+A_{%
\vec{k},\pm }^{out\dagger }(-\omega )e^{-i\phi _{\pm }(-\omega )}  \nonumber
\\
iY_{\pm } &=& A_{\vec{k},\pm }^{out}(\omega )e^{i\phi _{\pm }(\omega )}-A_{%
\vec{k},\pm }^{out\dagger }(-\omega )e^{-i\phi _{\pm }(-\omega )}
\label{rotate}
\end{eqnarray}%
The squeezing spectra \cite{book1} which measure the fluctuation of the
canonical position and momentum are defined by%
\begin{eqnarray}
S_{\vec{k}, X_{\pm }}(\omega ) &=&\left\langle X_{\pm }(\omega
)X_{\pm }(-\omega
)\right\rangle _{in},  \nonumber \\
S_{\vec{k}, Y_{\pm }}(\omega ) &=&\left\langle Y_{\pm }(\omega
)Y_{\pm }(-\omega )\right\rangle _{in}.  \label{S}
\end{eqnarray}%
where the $ \delta( \omega + \omega^{\prime} ) $ function was
omitted for notational simplicity, the in-state is the vacuum state
of the input field $ |BEC \rangle |0 \rangle $  shown in Fig.4.
Because the average $\left\langle X_{\pm }(\omega )\right\rangle
_{in}=\left\langle Y_{\pm }(\omega )\right\rangle _{in}=0$, then the
squeezing spectrums are $ S_{X_{\pm }}(\omega )= | \Delta X_{\pm
}(\omega) |^{2}$ and $ S_{Y_{\pm }}(\omega )=| \Delta Y_{\pm
}(\omega) |^{2}$. It can be shown that
\begin{eqnarray}
S_{\vec{k},X_{\pm }}(\omega ) &=&1+\left\langle :X_{\pm }(\omega )X_{\pm
}(-\omega ):\right\rangle _{in},  \nonumber \\
S_{\vec{k},Y_{\pm }}(\omega ) &=&1+\left\langle :Y_{\pm }(\omega )Y_{\pm
}(-\omega ):\right\rangle _{in}.
\end{eqnarray}%
where $:AB:$ denotes the normal order of the $A$ and $B$ with
respect to the $ A_{\vec{k},\pm }^{out}(\omega ) $, but the average
is taken with the incoming vacuum state.

For notational conveniences, we set $\phi _{-}(\omega )=\pi /2+\phi
_{+}(\omega )$ and just set $\phi _{+}(\omega )\equiv \phi (\omega )$. Then
we find $S_{X_{+}}(\omega )=S_{X_{-}}(\omega )=S_{X}(\omega )$ and $%
S_{Y_{+}}(\omega )=S_{Y_{-}}(\omega )=S_{Y}(\omega )$. The phase
$\phi (\omega )$ in the Fig.5 is chosen to achieve the largest
possible squeezing, namely, by setting $\partial S_{X}(\omega
)/\partial \omega =0$ which leads to:
\begin{equation}
\cos 2\phi (\omega )=\frac{\gamma _{\vec{k}}(\epsilon _{\vec{k}}+\bar{n}%
V_{d}(\vec{k}))}{\sqrt{\Omega ^{2}(\omega )+\gamma _{\vec{k}}^{2}E^{2}(\vec{k%
})+(\bar{n}V_{d}(\vec{k})\gamma _{\vec{k}})^{2}}},  \label{angle}
\end{equation}%
where $\Omega (\omega )=\omega ^{2}-E^{2}(\vec{k})+\gamma
_{\vec{k}}^{2}/4$, the $ \epsilon_{\vec{k}} $ and $ E(\vec{k}) $ are
related by Eqn.\ref{qspectrum}.

\begin{figure}[tbp]
\includegraphics[bb=61 150 561 664, width=6cm, clip]{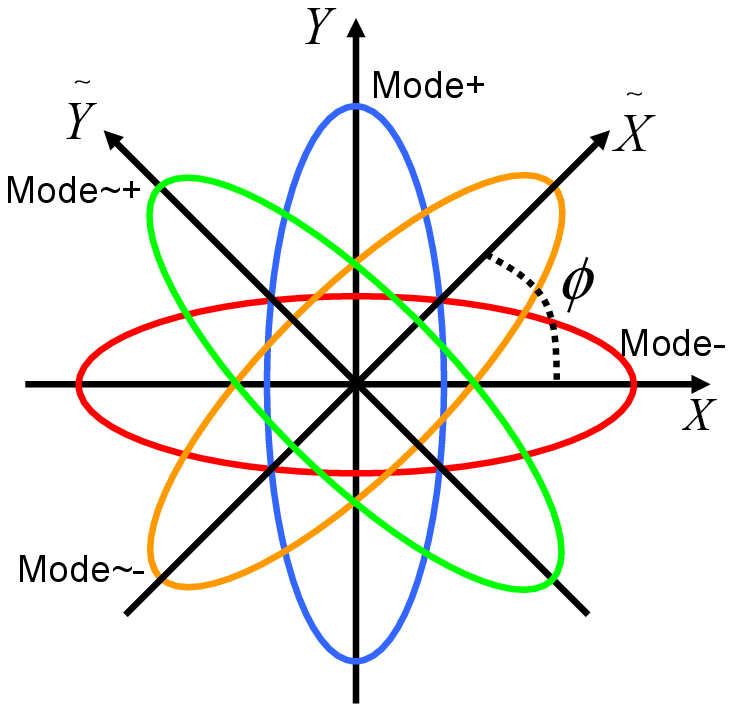}
\caption{The two mode squeezed state in the quadrature phase space.
Only when the photon frequency resonates with the quasi-particle
energy at $E(\vec{k} )
> \protect\gamma_{k}/2 $, the two mode squeezed states are mode + and mode
-. All the other cases have non-zero squeezing angles shown as mode
$\tilde{} \pm $. See also Fig.9. } \label{fig5}
\end{figure}

Substituting Eqns.\ref{bout},\ref{Aout} and \ref{rotate} into Eq. (\ref{S})
leads to%
\begin{eqnarray}
S_{X}(\omega ) &=&1-\frac{2\gamma _{\vec{k}}\bar{n}V_{d}(\vec{k})}{\mathcal{N%
}(\omega )+\gamma _{\vec{k}}\bar{n}V_{d}(\vec{k})}  \nonumber \\
S_{Y}(\omega ) &=&1+\frac{2\gamma _{\vec{k}}\bar{n}V_{d}(\vec{k})}{\mathcal{N%
}(\omega )-\gamma _{\vec{k}}\bar{n}V_{d}(\vec{k})} \label{squ}
\end{eqnarray}%
where $\mathcal{N}(\omega )=\sqrt{\Omega ^{2}(\omega )+\gamma _{\vec{k}%
}^{2}E^{2}(\vec{k})+(\bar{n}V_{d}(\vec{k})\gamma _{\vec{k}})^{2}}$

Obviously:
\begin{equation}
S_{X}(\omega )S_{Y}(\omega )=\Delta X_{\pm }\Delta Y_{\pm }=1
\label{squprod}
\end{equation}%
The results show that for a given in-plane momentum $\vec{k}$ and a
given  individual photon frequency $\omega $ with respect to the
chemical potential $\mu $, there always exists a two mode squeezing
state which can be decomposed into two squeezed states along two
normal angles: one
squeezed along the angle $\phi (\omega )$ and the other along the angle $%
\phi (\omega )+\pi /2$ in the quadrature phase space ($X,Y$) as
shown in Fig.5.

In the following, we discuss two cases $|\vec{k} |< k^{\ast} $ and
$|\vec{k} | > k^{\ast} $ respectively. For this purpose, we draw the
exciton energy $E( \vec{k} ) $ and the decay rate
$\gamma_{\vec{k}}/2 $ in the same plot in Fig.6. When $|\vec{k} |<
k^{\ast} $, the excitons decay very fast into photons, so they are
not well defined quasi-particles. However, when  $|\vec{k} | \gg
k^{\ast} $, the excitons decay into photons very slowly, so they are
well defined quasi-particles.

\begin{figure}
\includegraphics[width=7cm]{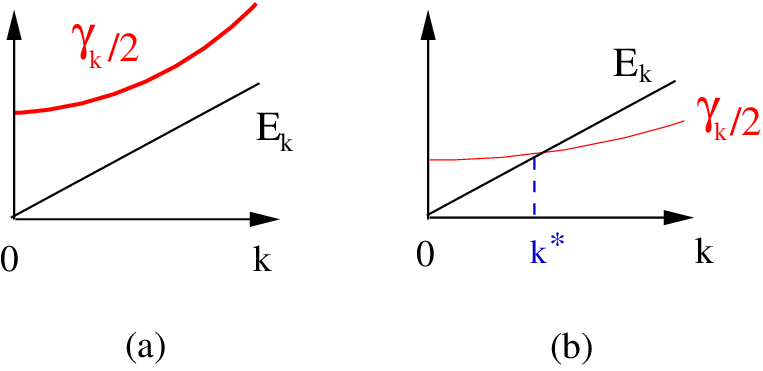}
\caption{The energy spectrum and the decay rate of the exciton
versus in-plane momentum $\vec{k}$. The direct exciton is always in
case (a), the quasi-particles in Fig.4 are not well defined in any
momentum scales. However, as argued in section IX,  due to its large
lifetime $ \tau_{ex} $, the indirect exciton is in case (b). When
$|\vec{k} |< k^{\ast} $, the quasi-particles in Fig.4 are not well
defined . However, when $|\vec{k} | \gg k^{\ast} $, they are well
defined quasi-particles. Compare with Fig.4 which is the case with $
\tau_{ex} \rightarrow \infty $. } \label{fig6}
\end{figure}

{\sl (1) Strong coupling case $|\vec{k}| < k^{\ast }$:
$E(\vec{k})<\gamma _{\vec{k}}/2$. }

From Eqn.\ref{squ}, we can see that the maximum squeezing happens at
$\omega _{\min }=0$ which means at $\omega _{k}=\mu $:
\begin{eqnarray}
S_{X}(\vec{k},\omega =0) &=&1-\frac{2\gamma _{\vec{k}}\bar{n}V_{d}(\vec{k})}{%
\mathcal{N}(0)+\bar{n}V_{d}(\vec{k})\gamma _{\vec{k}}}  \nonumber \\
\cos 2\phi (\vec{k},\omega =0) &=&\frac{\gamma _{\vec{k}}(\epsilon _{\vec{k}%
}+\bar{n}V_{d}(\vec{k}))}{\mathcal{N}(0)}  \label{S2}
\end{eqnarray}%
where $\mathcal{N}(0)\equiv \mathcal{N}(\omega =0)=\sqrt{[E^{2}(\vec{k}%
)+\gamma _{\vec{k}}^{2}/4]^{2}+(\bar{n}V_{d}(\vec{k})\gamma _{\vec{k}})^{2}}$%
which is defined below Eqn.\ref{squ}.  In sharp contrast to the weak
coupling case $ E(\vec{k})
> \gamma _{\vec{k}}/2$ to be discussed in the following, the resonance position $\omega
_{k}=\mu $ is independent of the value of $ \vec{k} $, this is
because the quasiparticle is not even well defined in the strong
coupling case. The $\omega $ dependence of $S_{X}(\omega )$ in
Eqn.\ref{squ} is drawn in Fig.7. The line width of the single peak
in Fig. 7 is:
\begin{equation}
\delta_{1}(\vec{k})=2\sqrt{E^{2}(\vec{k})-\frac{\gamma _{\vec{k}}^{2}}{4}+O_{%
\vec{k}}},  \label{width1}
\end{equation}%
where $
O_{\vec{k}}=\sqrt{4\mathcal{N}(0)[\mathcal{N}(0)+\bar{n}V_{d}(\vec{k})\gamma
_{\vec{k}}]-\gamma _{\vec{k}}^{2}E^{2}(\vec{k})} $.

\begin{figure}
\includegraphics[width=8cm]{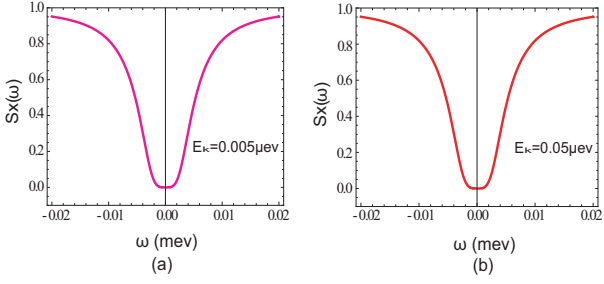}
\caption{The squeezing spectrum  at  a given in-plane momentum $
\vec{k} $ when  $ E(\vec{k}) < \gamma_{\vec{k}}/2 $. The $ n V_{d}(
\vec{k}) $ and $ \gamma_{\vec{k}}/2 $ are fixed at $ 50 \mu eV $ and
$ 0.1 \mu eV $ respectively as determined from experimental data in
section IX. They are also used in the following figures 9,10,12-16.
There exists only one minimal when the photon frequency equals to
the chemical potential with the width $ \delta_{1}( \vec{k}) $ given
by Eqn.\ref{width1}. Near the resonance, the squeezing ratio is so
close to  zero that it can not be distinguished in the figure. There
are very little difference between the two energies $
E(\vec{k})=0.005 \mu eV $ and $ E(\vec{k})=0.05 \mu eV $. This is
because the quasi-particles are not well defined in this case, the
squeezing spectrum is mainly determined by $ n V_{d}( \vec{k}) $ and
$ \gamma_{\vec{k}}/2 $. } \label{fig7}
\end{figure}

{\sl (2) Weak coupling case $ k > k^{\ast} $:  $ E(\vec{k})> \gamma
_{\vec{k}}/2$.}

As shown in Fig.6, in this case, $|\vec{k} | > k^{\ast} $. From Eqn.\ref{squ}%
, we can see that the maximum squeezing happens at the resonance frequency $%
\omega _{\min }=\pm \lbrack
E^{2}(\vec{k})-\gamma_{\vec{k}}^{2}/4]^{1/2}$. Recall that the
Fourier transformation of the Eq. (\ref{io}) gives $\omega
=\omega_{k}-\mu $, so the resonance condition becomes $\omega
_{k}=\mu \pm \lbrack E^{2}(\vec{k})-\gamma _{\vec{k}}^{2}/4]^{1/2}$
where
\begin{eqnarray}
S_{X}(\vec{k},\omega _{\min }) & = & ( \frac{ \epsilon_{\vec{k}}}{ E(\vec{k}%
) })^{2} = \frac{ \hbar
k^{2} }{ \hbar^{2} k^{2} + 4 M \bar{n}V_{d}(\vec{k}) }  \nonumber \\
\cos 2 \phi(\vec{k},\omega _{\min }) & = & 1  \label{S1}
\end{eqnarray}
In this case, $\phi (\vec{k}, \omega _{\min })=0$, so the two
squeezing modes are mode + and mode - respectively shown in Fig.5.
In sharp contrast to the strong coupling case discussed above,   the
resonance positions depend on the value of $ \vec{k} $, this is
because the quasiparticle is  well defined in the weak coupling
case. The squeezing ratio is independent of $ \gamma_{\vec{k}} $ at
the resonances ! Of course, away from the resonances, it will always
depends on $ \gamma_{\vec{k}} $. From Eqn.\ref{S1}, we can see that
increasing the exciton mass, the density and the exciton dipole
interaction will all benefit the squeezing.


The $\omega $ dependence of $S_{X}(\omega )$ in Eqn.\ref{squ} is
drawn in
Fig.8. When $E(\vec{k})>(Q_{\vec{k}}+\sqrt{1+Q_{\vec{k}}})\gamma _{\vec{k}%
}/2 $, the line width of the each peak in Fig.8 is
\begin{eqnarray}
\delta_{2}(\vec{k}) &=&\sqrt{E^{2}(\vec{k})-\frac{\gamma _{\vec{k}}^{2}}{4}+\gamma _{\vec{%
k}}Q_{\vec{k}}E(\vec{k})}   \\  \nonumber
&&-\sqrt{E^{2}(\vec{k})-\frac{\gamma _{\vec{k}}^{2}}{4}-\gamma _{\vec{k}}Q_{%
\vec{k}}E(\vec{k})}
\label{width2}
\end{eqnarray}%
where $ Q_{\vec{k}}=\sqrt{3+\frac{4\bar{n}V_{d}(\vec{k})}{\epsilon
_{\vec{k}}}} $. It is easy to see that $ \delta_{2} \sim
\gamma_{\vec{k}} Q_{\vec{k}} $ which is equal to the exciton decay
rate $ \gamma_{\vec{k}} $ multiplied by a prefactor $ Q_{\vec{k}} $.

When $E(\vec{k})<(Q_{\vec{k}}+\sqrt{1+Q_{\vec{k}}})\gamma
_{\vec{k}}/2$, the two peaks are too close to be distinguished.

\begin{figure}
\includegraphics[width=8cm]{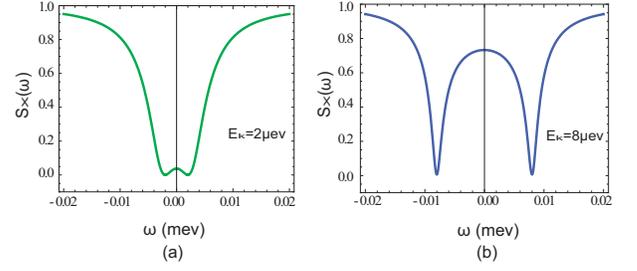}
\caption{ The squeezing spectrum  at  a given in-plane momentum $
\vec{k} $ when  $ E(\vec{k}) > \gamma _{\vec{k}}/2 $. There exist
two minima in the spectrum when the photon frequency resonate with
the well defined quasi-particles. Near the resonance, the squeezing
ratio is so close to zero that it can not be distinguished in the
figure. When $E(\vec{k})=2 \mu eV $, the two peaks are still not
clearly separated. When $E(\vec{k})=8 \mu eV \gg \gamma
_{\vec{k}}/2$, the quasi-particles are well defined which lead to
the two well defined resonances with width $ \delta_{2}( \vec{k} ) $
in Eqn.\ref{width2}. } \label{fig8}
\end{figure}

In short, for a given in-plane momentum, there always exists a two
mode squeezed state. When
$E(\vec{k})<\gamma_{\vec{k}}/2$, the squeezing spectrum reaches its minimum Eq. \ref{S2} at $%
\omega _{k}=\mu $ and the squeezed angle is always non zero $\phi
(\omega )\neq 0$. On the other hand, when $E(\vec{k})>\gamma
_{\vec{k}}/2$, the squeezing spectrum reaches its minimum Eq.
\ref{S1} at $\omega _{k}=\mu \pm \lbrack E^{2}(\vec{k})-\gamma
_{\vec{k}}^{2}/4]^{1/2} $ and the squeezed angle $\phi (\omega
_{\min })=0$.  In sharp contrast to the widths in the ARPS and EDC
in Fig.11,13 to be discussed in the Sec.VII which depend only on $
\gamma _{\vec{k}} $, the two widths in Eqn.\ref{width1} and
\ref{width2} in the squeezing spectra  also depend on the
interaction! The angle dependence of Eqn.\ref{angle}  in both the
strong coupling $ E(\vec{k})< \gamma _{\vec{k}}/2$  and the weak
coupling $E(\vec{k})>\gamma _{\vec{k}}/2$  cases are drawn in the
same plot Fig.9 for comparison. From Eq. \ref{S2} which is valid at
$E(\vec{k})<\gamma _{\vec{k}}/2$
( $|\vec{k}| < k^{\ast }$ ) and  Eq. \ref{S1} which is valid at $E(\vec{k})>\gamma _{\vec{k}}/2$ ( $|%
\vec{k}| > k^{\ast }$ ) at the resonance, we can find that
the squeezing ratio and the angle dependence at the resonance on the whole in-plane momentum $%
\vec{k}$ regime.

\begin{figure}
\includegraphics[width=8 cm]{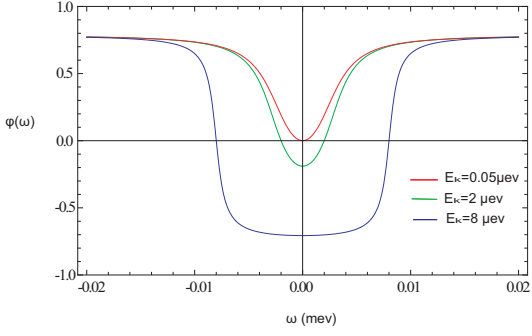}
\caption{ The squeezing angle dependence on the frequency when
$E(\vec{k})< \gamma _{\vec{k}}/2$ and $E(\vec{k})> \gamma
_{\vec{k}}/2$.  When $E(\vec{k})< \gamma _{\vec{k}}/2$ the squeezing
angle is always non-zero. Near the resonance, the angle is so close
to zero that it can not be distinguished in the figure. Only when
$E(\vec{k})> \gamma _{\vec{k}}/2$ and the photon frequency resonate
with the well defined quasi-particles, the squeezing angle is zero.
Away from the resonance, the angle becomes negative.}\label{fig9}
\end{figure}

\textsl{(3) Phase sensitive Homodyne  measurement to measure the
squeezing spectrum  and the rotating phase  $ \phi(\omega) $ }

   Usual measurements are just intensity  measurement such as power spectrum experiment and
   intensity-intensity correlation measurement such as HBT experiment \cite{book1,book2}, so contain no phase information.
   Detection of squeezed states, on
   the other hand, requires a phase sensitive scheme that measures
   the variance of a quadrature of the photon field. This can be achieved
   the phase sensitive homodyne detection.  The experimental
   set-up of this kind of experiment was explained in detail in
   \cite{book1,book2}, here we just briefly  explain the main
   points of a {\sl single mode} phase sensitive homodyne detection by a schematic Fig.10.
   This figure need to be connected with the homodyne outputs of
   the Fig.16 to detect the two modes squeezing spectrum
   Eqn.\ref{squ} and squeezing angle Eqn.\ref{angle}.
   It is essentially a phase interference experiment between the input
   light beam  A and a local oscillator (LO) beam  B which is used as a reference beam.  Both input beam A and  LO beam B are
   incident on a beam splitter and  are reflected and
   transmitted, there is a  $ \pi/2 $ phase shift between the reflected and the transmitted beam.
   so there are two beams C and D coming out the splitter, both C and D are
   linear combination of A and B. If one fixed the LO
   beam B to be a strong coherent field with phase $ \phi $, a
   balanced detection is to use a  50/50 beam splitter, the output
   signal detected by the coincidence measurement in the Fig.10 is taken to be the difference between the counting of C photons and
   that of D photons, so it is just the interference
   between the quadrature of the input beam A and the strong LO beam B
   subject to a rotation  by angle $ \phi+\pi/2 $. The variance of the output can also be
   measured which is just the squeezing spectrum Eqn.\ref{S}.  By
   tuning the angle $ \phi $, both $ X $ quadrature and $ Y $ quadrature or its any linear
   combination quadrature of the input beam can be measured. Then the rotated phase $ \phi
   $ in Eqn.\ref{angle} shown in Fig.5 and drawn in Fig.9 is just this phase $ \phi $ of the local
   oscillator, so they are completely experimental measurable
   quantities in the phase sensitive homodyne experiments.
\begin{figure}
\includegraphics[width=6cm]{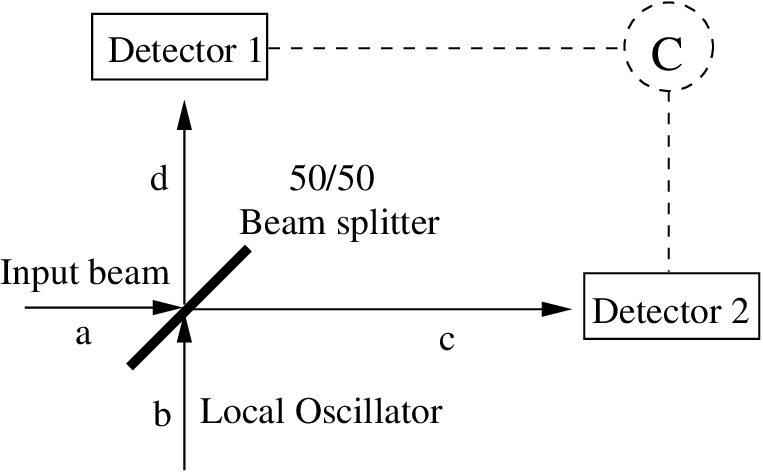}
\caption{ The balanced homodyne experiment to measure a single mode
squeezing spectrum. } \label{fig10}
\end{figure}

\section{ One photon correlation function, Power spectrum and
Macroscopic super-radiance}

The one photon correlation function of the output field is $G_{\pm
}(\tau
)=\left\langle a_{\pm \vec{k}}^{out\dagger }(t+\tau )a_{\pm \vec{k}%
}^{out}(t)\right\rangle _{in}$ and the angle resolved power spectrum (ARPS) of the output field is $%
S_{\pm }(\vec{k}, \omega )=\int_{-\infty }^{+\infty }d\tau
e^{-i\omega \tau }G_{\pm
}(\tau )$. The normalized first order correlation function is defined by $%
g_{\pm }(\tau )=G_{\pm }(\tau )/G_{\pm }(0)$. In the following, we
will evaluate these quantities respectively.

{\sl (1) The angle resolved power spectrum (ARPS) }

 By inserting
Eqn.\ref{bout}, one obtains the photon number spectrum $S_{\pm
}(\vec{k}, \omega )=S_{1}(\vec{k}, \omega )$ and
\begin{eqnarray}
S_{1}(\vec{k},\omega ) & = & \frac{1}{4} ( S_{X}( \omega )+ S_{Y}(
\omega )
-2 )           \nonumber \\
  & = & \frac{ \gamma _{\vec{k}}^{2} \bar{n}^{2}V_{d}^{2}(\vec{k})}{
\Omega ^{2}(\omega )+\gamma _{\vec{k}}^{2}E^{2}(\vec{k})}.
\label{power}
\end{eqnarray}%
where $\Omega (\omega )=\omega
^{2}-E^{2}(\vec{k})+\gamma_{\vec{k}}^{2}/4$.
  From Eqn.\ref{squprod}, one can see that $ S_{X}( \omega )+ S_{Y}(
\omega )  \geq 1 $, so the more squeezing, the stronger the power
spectrum. If there is no squeezing $ S_{X}( \omega )= S_{Y}( \omega
) = 1 $, then there is no power emitted, this is just the input
vacuum state in the Fig.4 which is a coherent state itself.
The angle resolved power spectra (ARPS) with different $E(\vec{k})$ and $\gamma _{\vec{%
k}}$ are shown in Fig.11. The total ARPS is the sum of the
condensate and the quasi-particles: $ S(\vec{k}, \omega )=
S_{0}(\vec{k}, \omega )+S_{1}(\vec{k}, \omega ) $.

In the strong coupling case $ k< k^{*}, E(\vec{k})<\gamma
_{\vec{k}}/2$, $ S_{1}(\vec{k},\omega )$ reaches the maximum $\gamma _{\vec{k}}^{2}\bar{n}%
^{2}V_{d}^{2}(\vec{k})/[\gamma _{\vec{k}}^{2}/4+E^{2}(\vec{k})]^{2}  $ at $%
\omega _{k}=\mu $. As $ \vec{k} \rightarrow 0 $, $
   E(\vec{k}) = u |\vec{k}| \rightarrow 0 $, then $
   S_{1}(\vec{k},\omega ) \rightarrow
   \frac{ \gamma _{\vec{k}}^{2} \bar{n}^{2}V_{d}^{2}(\vec{k})}
   { (\omega ^{2}+ \gamma _{\vec{k}}^{2}/4 )^{2}} $, so the curve has a half
   width $ \sim \hbar \gamma_{0} \sim 10^{-4} meV $. This is
   expected, because the quasiparticles are not well defined
   with the decay rate $ \gamma_{0} $ much larger than its energy $  E( \vec{k})$.

In the weak coupling case $E(\vec{k})>\gamma _{\vec{k}}/2$, at the
two resonance
frequencies $\omega _{k}=\mu \pm \lbrack E^{2}(\vec{k})-\gamma _{\vec{k}%
}^{2}/4]^{1/2}$, $S_{1}(\vec{k}, \omega )$ reaches the maximum $\bar{n}^{2}V_{d}^{2}(%
\vec{k})/E^{2}(\vec{k})$ which only depends on the exciton density,
the dipole-dipole interaction and the quasi-particle spectrum, but
independent of $ \gamma _{\vec{k}} $ ! It can be shown that when
$E(\vec{k}) \gg \gamma _{\vec{k}}/2$, the width of the two peaks at
the two resonance frequencies is  $ \sim \gamma _{\vec{k}} $, this
is expected, because the quasi-particle is well defined with energy
$ \lbrack E^{2}(\vec{k})-\gamma _{\vec{k}}^{2}/4]^{1/2}$ and the
 half-width $ \gamma_{\vec{k} }$. In sharp contrast to the two widths in Eqn.\ref{width1} and
\ref{width2} in the squeezing spectra  which depend on both the
interaction and $ \gamma_{\vec{k}} $, the widths in the ARPS and EDC
in Fig.11,13  depend only on $ \gamma _{\vec{k}} $.

\begin{figure}
\includegraphics[width=6.5cm,height=5cm]{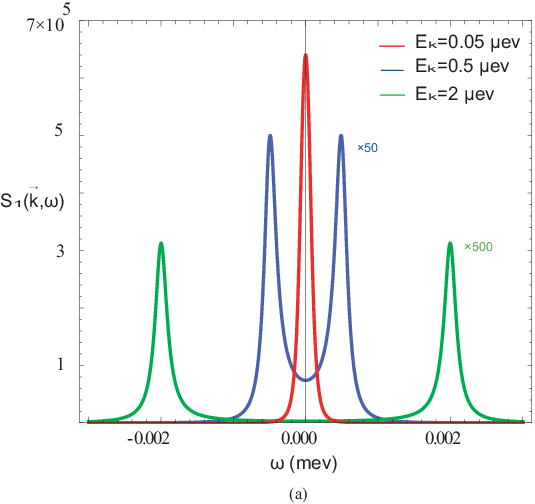}
\caption{ The angle resolved power spectrum (ARPS)
of the emitted photon with in-plane momentum $ \vec{k} $.  When
$E(\vec{k})\leq \hbar \gamma_{\vec{k}}/2$, there is only one peak in
the power spectrum with the width $ \gamma_{\vec{k}}/2 $.  When
$E(\vec{k})>\hbar \gamma_{\vec{k}}/2$, there are two peaks in the
power spectrum at the two resonance photon frequencies $ \omega
_{k}\simeq \mu \pm E(\vec{k})$ also with width $ \gamma_{\vec{k}}/2
$. The $ S_{1}(\vec{k},\omega ) $ at  $ E( \vec{k})=0.5 \mu eV $ and
$ E( \vec{k})=2 \mu eV $ are multiplied by $ 50 $ and $ 500 $ in
order to be seen in the figure. Compared to the squeezing spectrum
in Fig.9, one can see the ARPS can distinguish the two
quasi-particle peaks clearly even at $E(\vec{k}) = 0.5 \mu eV $.
While the squeezing spectrum in Fig.9 can not distinguish the two
quasi-particle peaks even at $E(\vec{k}) = 2 \mu eV $, so the ARPS
is a much sensitive probe of the quasi-particle spectrum than the
squeezing spectrum. } \label{fig11}
\end{figure}

 {\sl (2) Momentum Distribution Curve (MDC) }

  The power spectrum at a given in-plane momentum $ \vec{k} $ is
  $ S_{1}(\vec{k} ) = \sum_{k_z} S_{1}(\vec{k}, \omega ) =
  \int d \omega_{k} D_{\vec{k}}(\omega_k )S_{1}(\vec{k}, \omega )  $ which is
  nothing but the Momentum Distribution Curve (MDC) \cite{myprl}:
\begin{equation}
   S_{1}(\vec{k} )= \frac{ D_{\vec{k}}(\mu )
   \bar{n}^{2}V_{d}^{2}(\vec{k})\gamma_{\vec{k}}}
   {2[ E^{2}(\vec{k} ) + (\frac{\gamma _{\vec{k}}}{2})^{2}]  }  \propto L_{z}
\label{mdc}
\end{equation}
  As to be explained in section IX and shown in the Fig.6, the crossing point is at $ k^{\ast} \sim
  10^{-2} cm^{-1} $. From Eqn.\ref{npht}, one can see the
  condensate contribution at $ \vec{k}=0 $ is $ N_{ph}/L_{z} = N
  \gamma_{0}/v_{g} \propto N $, while the  contribution from the
  quasi-particle is $ S_{1}(\vec{k} \rightarrow 0 )/L_z = \frac{ 2 n^{2} V^{2}_{d}(0) }{ v_{g} \gamma_{0} }
  \propto n^{2}/\gamma_{0} $.
  So the MDC is a Lorentian with the half width
  at $ k^{\ast} $ as shown in the Fig.12.  So the $ k^{\ast} $ has a clear
  physical meaning as the half width of the MDC and is an
  experimentally measurable quantity.

  In fact, we can also calculate the one photon correlation function:
\begin{equation}
  G(r) \sim \int \frac{ d^{2} \vec{k} }{ ( 2 \pi )^{2} } \frac{ e^{i
  \vec{k} \cdot \vec{r}} } { k^{2}+ k^{*2}} \sim e^{-k^{*} r }
\label{coh}
\end{equation}
   where we can identify the coherence length $ \xi \sim1/k^{*} \sim
   40  \mu m $. This coherence length has been measured in
   \cite{coherence} and will be discussed in detail in section IX.

\begin{figure}
\includegraphics[width=3.5cm]{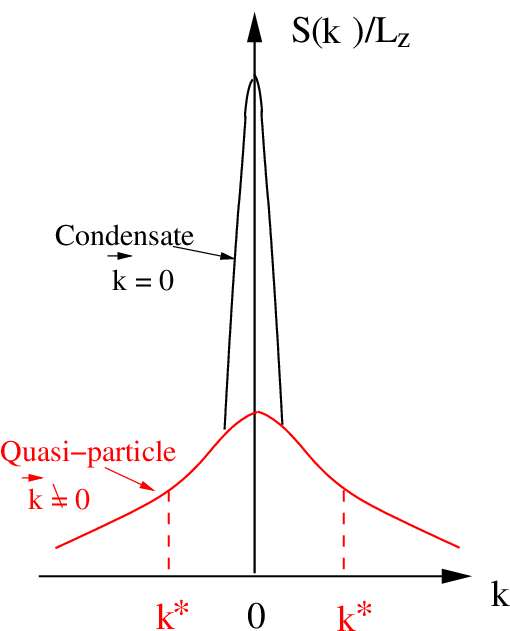}
\caption{ The zero temperature Momentum Distribution Curve (MDC) has
a bi-model structure consisting of a sharp peak  $ S( \vec{k}=0
)/L_z=  N \gamma_{0}/v_g $ due to the condensate at $ \vec{k}=0 $
superposing on a Lorentzian peak with a half width $ k^{\ast} \sim
10^{2} cm^{-1} $ due to quasi-particle excitations at $ \vec{k} \neq
0 $. The coherence $ \xi \sim 1/k^{*} $ has been measured in
\cite{coherence}. }\label{fig12}
\end{figure}

 {\sl (3) Energy Distribution Curve (EDC) }

  The power spectrum at a given energy $ \omega $
  is $ S_{1}(\omega ) = \sum_{\vec{k}} S_{1}(\vec{k}, \omega ) $ which is
  nothing but the Energy Distribution Curve (EDC)\cite{myprl}:
\begin{eqnarray}
   S_{1}( \omega ) & = &  N \bar{n} \times \int \frac{ d^{2} \vec{k}}{(2 \pi)^{2}}
\frac{ \gamma _{\vec{k}}^{2} V_{d}^{2}(\vec{k})}{ \Omega ^{2}(\omega
)+\gamma _{\vec{k}}^{2}E^{2}(\vec{k})}   \nonumber   \\
    & = & \frac{ N \bar{n} V^{2}_{d}( \vec{k}=0) }{ 4 \pi u^{2} } f(
    \frac{ |\omega|}{ \gamma_{\vec{k}=0 } }) \propto  N \bar{n}
\label{edc}
\end{eqnarray}
   where $ f(x)= \frac{1}{x} [ \frac{\pi}{2} -arctg \frac{ 1/4-x^{2}
   }{x} ] $ where $ - \pi/2 < arctg y < \pi/2  $.
   Because $ f(0) =4 $, then $ S_{1} ( \omega =0 )/N = \frac{ \bar{n} V^{2}_{d}( \vec{k}=0 ) }{  \pi u^{2}
   } $. The $ S_{1}( \omega ) $ is shown in Fig.13a.

   The $ S_{1}( \omega ) $ is the sum over all the angle resolved power spectrum curves in Fig.11 at the fixed energy
   $ \omega= \omega_{k}- \mu $. As $ \vec{k} \rightarrow 0 $, $
   E(\vec{k}) = u |\vec{k}| \rightarrow 0 $, then $
   S_{1}(\vec{k},\omega ) \rightarrow
   \frac{ \gamma _{\vec{k}}^{2} \bar{n}^{2}V_{d}^{2}(\vec{k})}
   { (\omega ^{2}+ \gamma _{\vec{k}}^{2}/4 )^{2}} $, so the curve has a half
   width $ \sim \hbar \gamma_{0} \sim 10^{-4} meV $.
   As $ |\vec{k} |
   $ increases to $ k^{*} \sim 10^{-2} cm^{-1} $, the curve starts
   to split into two peaks as shown in the Fig.11, then when $ |\vec{k}
   | \rightarrow |\vec{k}_{max}  |/2 $, the two peaks stand for the two well defined
   quasi-particle excitations, the splitting of the two
   peaks reaches $ \sim   u |\vec{k}_{max}  |\sim 0.1 meV $.
   So the EDC
   curve will simply smear out all the fine structures of the angle
   resolved power spectrum in Fig.11 and finally end up with an envelop curve with a half width $
   w_{e} \sim  \gamma_{\vec{k}} $ as shown in the Fig.13.
   From Eqn.\ref{nph1}, one can see the
   condensate contribution at $ \vec{k}=0 $ is $ n_{ \omega_{k_z}}/N =
   \gamma_{0} \delta ( \omega_{k_z} - \mu )  $, while the  contribution from the
   quasi-particle $ S_{1} ( \omega =0 )/N = \frac{ \bar{n} V^{2}_{d}( \vec{k} ) }{  \pi u^{2}
   } $ as shown in the Fig.13a.
   As to be explained in
   section IX, the EDC is what experiments in \cite{butov,snoke}
   measured at various $ P_{ex}, V_{g} $ and $ T $.

\begin{figure}
\includegraphics[width=3.5cm]{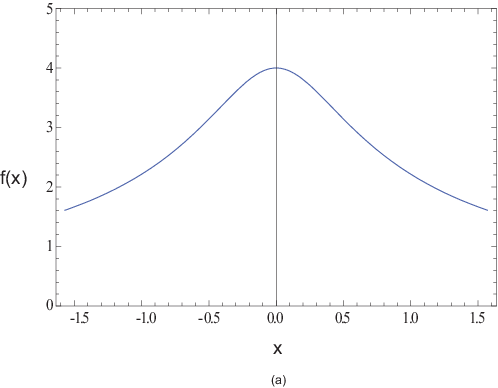}
\hspace{0.5cm}
\includegraphics[width=3.5cm]{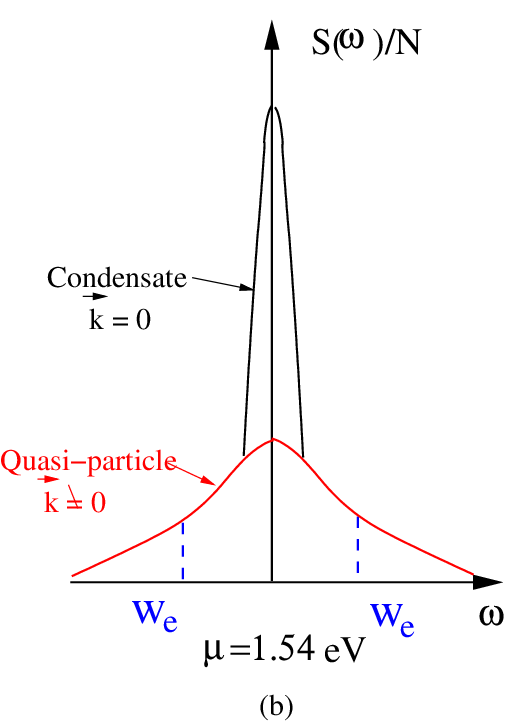}
\caption{ (a) The Energy Distribution Curve ( EDC ) from the
quasi-particle. (b) The zero temperature EDC has a bi-model
structure consisting of a sharp $ \delta $ function peak $ S( \omega
)/N = \gamma_{0} \delta( \omega - \mu ) $ due to the condensate at $
\vec{k}=0 $ superposing on a Lorentzian peak with the half envelop
width $ w_{e} \sim \gamma_{\vec{k}}  \sim 0.1  \mu eV $ due to
quasi-particle excitations at $ \vec{k} \neq 0 $.}\label{fig13}
\end{figure}

 {\sl (4) The total radiation rate from the quasi-particles }

   The total number of photons emitted from the quasi-particles is:
\begin{equation}
    S_{1}= \sum_{k_{z}} \sum_{\vec{k}}S_{1}(\vec{k}, \omega )
   \sim N \times L_{z} \bar{n}
\label{total}
\end{equation}
   which is  proportional to the total
   normalization volume of the system as it is expected. This can
   also be used as a self-consistency check on our results achieved
   on the quasi-particle part at $ \vec{k} \neq 0 $. As shown in
   the section IV, this self-consistency check played very important roles on the condensate
   part $ \vec{k}=0 $. In fact, if taking Eqns.\ref{edc} and
   \ref{mdc} at face value, then the total number of photons
   Eqn.\ref{total} diverge, but this should not cause any concern,
   because both Eqns.\ref{edc} and
   \ref{mdc} only hold at small momentum $ k < k_{max} $ and $
   \omega < u k_{max} $  respectively.

   The radiation rate along a given direction $ ( \vec{k}, k_z ) $ is:
\begin{equation}
  P^{ra}_{1}(\vec{k}, k_z ) = \frac{  S_{1}( \vec{k}, \omega ) }{ L^{2} L_z }
  \omega_{k} v_{g} \times L^{2}
\label{qradiation}
\end{equation}
   which vanishes in the limit $ L_{z} \rightarrow \infty $ as expected.
   Then the radiation rate at a given in-plane momentum $
   P^{ra}_{1}(\vec{k} ) = v_{g} \mu S_{1}( \vec{k})/L_{z} $
   and the radiation rate at a given energy $
   P^{ra}_{1}(\omega ) = v_{g} \mu S_{1}( \omega )/L_{z} $
   where we have used the fact that $ S_{1}( \vec{k}, \omega ) $
   is a even function of $ \omega=\omega_{k}- \mu $.

   The total radiation rate from the quasi-particles is:
\begin{equation}
  P^{ra}_{1} =  \sum_{k_{z}} \sum_{\vec{k}} P_{rd}(\vec{k}, k_z )
  = v_{g} \mu  S_{1} /L_{z} \propto v_{g} \mu N \bar{n}
\end{equation}
  where again $ k_{z} $ is sum over both the upper and the lower
  space in the Fig.1b. $ P^{ra}_{1} $ is also $ \propto \mu N  $ just as the radiation rate $ P^{ra}_{0} $ from the
  condensate. Because of the weak dipole-dipole interaction, the quantum depletion is small, so
  $ P^{ra}_{1} $  is still much smaller than $ P^{ra}_{0} $ in Eqn.\ref{power0}.
  Furthermore $ P^{ra}_{1} $ is spread over all the possible angles,
  while $ P^{ra}_{0} $ is focused along a single direction.

{\sl (5) One photon correlation functions}

By using the Fourier transformation to $S_{1}(\vec{k}, \omega )$, one can get $%
G_{\pm }(\vec{k}, \tau )=G_{1}(\vec{k}, \tau )$
\begin{equation}
G_{1}(\vec{k}, \tau )=\frac{\bar{n}^{2}V_{d}^{2}(k)\gamma
_{k}}{4E(k)}[\frac{ e^{i(E(k)+i\frac{\gamma _{k}}{2})\tau
}}{E(k)+i\frac{\gamma _{k}}{2}}+h.c.],
\end{equation}

The normalized first order correlation function $g_{\pm }(\vec{k},
\tau )=g_{1}(\vec{k}, \tau )$, where
\begin{equation}
g_{1}(\vec{k}, \tau )=e^{-\frac{\gamma _{\vec{k}}}{2}\tau }[\cos
(E(\vec{k})\tau )+ \frac{\gamma _{\vec{k}}}{2E(\vec{k})}\sin
(E(\vec{k})\tau )] \label{g1}
\end{equation}
It turns out that the first order correlation function is
independent of the relation between $E(\vec{k})$ and $\gamma
_{\vec{k}}/2$ and is shown in Fig.14. The $ G_{1} $ was measured in
the EHBL in \cite{coherence} at $ T=1.6 K $ and in exciton polariton
in \cite{exp2} at $ T=4 K $. The effects of finite temperature and
trap potential must be considered before comparing our theoretical
results with the experimental data.

\begin{figure}
\includegraphics[width=7cm]{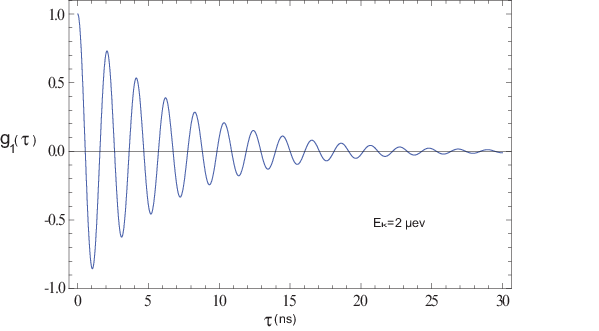}
\caption{ The one photon correlation function }\label{fig14}
\end{figure}

{\sl (6) The quasi-particle spectrum in a non-equilibrium stationary
exciton superfluid }

  It is important to compare the excitation spectrum in Fig.4,
  Fig.6a and Fig.6b. Fig.4 is the well know quasi-particle
  excitations in an equilibrium superfluid. They are well defined
  quasi-particles with infinite lifetime. However, the quasi-particles in Fig.6a
  are not well defined in any length scales, because the decay rate
  is always much larger than the energy. Fig.6(b) is between the two
  extreme cases. When $ k < k^{*} $, the quasi-particle is not well
  defined, the ARPS is centered around $ \omega_{k}= \mu $ with the
  width $ \gamma_{k} $. The MDC has large values at $ k < k^{*} $.
  The EDC has large values at $ \omega < \gamma_{k} $.
  When $ k > k^{*} $, the quasi-particles is well
  defined,  the ARPS has two well defined quasi-particles peaks at
  $ \omega_{k}=  \mu \pm \lbrack E^{2}(\vec{k})-\gamma _{\vec{k}}^{2}/4]^{1/2}$
  with the width $ \gamma_{k} $. The MDC has very small values at $ k > k^{*} $.
  The EDC has very small values at $ \omega > \gamma_{k} $.
  So in the long wavelength  $ r > \xi \sim 1/k^{*} $ ( or small momentum
  $ k < k^{* } $ ) limit and long time $ \tau > \tau_{ex} \sim 1/\gamma_{k} $
  (or low energy limit $ \omega < \gamma_{k} $ ) limit, there is
  not a well defined superfluid  which is consistent with the
  results achived in \cite{kohn}. However, in the distance sacle $ r < \xi \sim 1/k^{*} $ ( or momentum
  $ k > k^{* } $ ) limit and the time scale $ \tau < \tau_{ex} \sim 1/\gamma_{k} $
  (or  energy scale $ \omega > \gamma_{k} $ ), there is
  still defined superfluid and associated quasi-particle excitations.
  This is the main difference and analogy between the equilibrium
  superfluid in Fig.4 and the non-equilibrium steady state
  superfluid  in Fig.6b. Very recently, the elementary excitation spectrum of
  exciton-polariton inside a micro-cavity was measured \cite{expp} and
  was found to be very similar to that in a helium 4 superfluid shown
  in Fig.4 except in a small regime near $ k=0 $. We believe this
  observation is precisely due to the excitation spectrum in a
  non-equilibrium stationary superfluid shown in Fig.6b.

{\sl (7) The Superradiance from the quasi-particles }

 Note that the angle resolved power spectrum, the MDC and
EDC in Eqns.\ref{power} \ref{mdc}, \ref{edc} are all proportional to
$ N^{2} $ instead of $ N $. It is the characteristic of
super-radiance in a macroscopic system. This should not be too
surprising, because the excitonic superfluid is a macroscopic
quantum coherence phenomena, so it is natural to lead to macroscopic
superradiance. As $ k \rightarrow 0 $ in the Fig.6,  $
 S_{1}( \vec{k} ) \sim \frac{\bar{n}^{2}
V^{2}_{d}(k)}{\gamma_{\vec{k}}} \sim \bar{n}^{2}/\gamma_{\vec{k}}  $
where $ \gamma_{k} $ appears in the denominator in the strong
coupling case, so the macroscopic superradiance can only be achieved
by the non-perturbative calculations presented in this section, but
can not be derived by any finite order perturbative calculations
presented in the appendix D.

In conventional quantum optics, $ N $ two level atoms interacting
with a single ( or multi-) photon mode(s) inside a cavity. If the $
N $ static atoms are confined into a small volume $ V $  inside the
cavity which is much smaller than the wavelength of the photon mode,
then the interaction between all the $ N $ atoms and the photon mode
can be taken as the same constant $ \lambda $, then when half of the
atoms are in the excited level, the radiation intensity from the $ N
$ atoms is proportional to $ N^{2} $ instead of just $ N $ during
the time interval $ \sim 1/N $, so the total power emitted during
this time period is $ N^{2} \times 1/N \sim N $ as required by the
energy conservation. This is due to the cooperative effects of the $
N $ atoms which is due to the fact that the $ N $ atoms, being
interacting with the same photon field, so can not be treated as
independent $ N $ atoms. This is called non-equilibrium
superradiance first studied by Dicke \cite{super}. Generalizing the
Dicke model to a stationary state inside a cavity was studied in
\cite{dick1,dick2}. It was found that there is a second order phase
transition driving the coupling constant $ \lambda $: when $ \lambda
< \lambda_{c} $, the system is in a normal phase, when $ \lambda >
\lambda_{c} $, the system is in a superradiative phase. It was also
pointed out in \cite{dick2} that it is very unrealistic to realize
Dicke model in the thermodynamic limit $ N, V \rightarrow \infty $,
but keep $ N/V $ to be finite, because it is essentially impossible
to make $ V $ still smaller than the wavelength of the photon in the
thermodynamic limit. So the superradiance is essentially an effect
for finite number of static atoms confined into a small volume.

The superradiance from the ESF has completely different mechanism:
(1) the size of the sample is much larger than the wavelength of the
photon field (2) The photons field is a continuum of photons with
different $ k_z $ at a given $ \vec{k} $ in Eqn.\ref{chem}, so
acting as a reservoir to the excitons with in-plane momentum $
\vec{k} $ (3) all the excitons are always in motion, in fact, moving
in a coherent fashion. So all these conditions violate the
conditions to achieve the superradiance in conventional quantum
optics. So the collective radiation from the ESF is due to the
macroscopic coherence of the exciton superfluid itself which is, in
turn, due to the dipole-dipole repulsion.

\section{ Two photon correlation functions and photon statistics}

The quantum statistic properties of emitted photons can be extracted
from two photon correlation functions. The normalized second order
correlation functions of the output field for the two modes at $
\vec{k} $ and $ -\vec{k} $ are
\begin{equation}
g_{2}^{(\vec{k})}(\tau )=\frac{\left\langle a_{\vec{k}}^{out\dagger }(t)a_{%
\vec{k}}^{out\dagger }(t+\tau )a_{\vec{k}}^{out}(t+\tau )a_{\vec{k}%
}^{out}(t)\right\rangle_{in} }{\left\vert G_{1}(0)\right\vert ^{2}}
\label{g2}
\end{equation}%
and%
\begin{equation}
g_{2}^{(\pm \vec{k} )}(\tau )=\frac{\left\langle
a_{\vec{k}}^{out\dagger }(t+\tau
)a_{\vec{k}}^{out}(t+\tau )a_{-\vec{k}}^{out\dagger }(t)a_{-\vec{k}%
}^{out}(t)\right\rangle_{in} }{\left\vert G_{1}(0)\right\vert ^{2}}
\label{g2pm}
\end{equation}%
The second order correlation function $g_{2}^{(\pm \vec{k} )}(\tau
)$ determines the probability of detecting $n_{-\vec{k} }$ photons
with momentum $-\vec{k} $ at time $t$ and detecting $n_{\vec{k}}$
photons with momentum $ \vec{k} $ at time $t+\tau $. Just like the
one photon correlation function in Eqn.\ref{g1}, it turns out that
the second correlation functions are also independent of the
relation between $E(\vec{k})$ and $\gamma _{\vec{k}}/2$ and are
shown in Fig.15. The Wick theorem \cite{foot} gives
$g_{2}^{(\vec{k})}(\tau )=1+\left\vert g_{1}(\tau )\right\vert ^{2}$
where $ g_{1}(\tau ) $ is given by Eqn.\ref{g1}. Similarly, it can
be shown that $g_{2}^{(\pm \vec{k} )}(\tau )=1+\left\vert f_{1}(\tau
)\right\vert ^{2}$ where
\begin{eqnarray}
f_{1}(\tau ) &=&-\frac{E^{2}(\vec{k})+\frac{\gamma
_{\vec{k}}^{2}}{4}}{\bar{n}V_{d}(\vec{k})}e^{-\frac{\gamma _{\vec{k}}}{2}\tau }  \nonumber \\
& \times &[u_{\vec{k}}^{2}\frac{e^{-iE(\vec{k})\tau
}}{E(\vec{k})-i\frac{\gamma
_{\vec{k}}}{2}}+v_{\vec{k}}^{2}\frac{e^{iE(\vec{k})\tau
}}{E(\vec{k})+i\frac{ \gamma _{\vec{k}}}{2}}].
\end{eqnarray}
Then the normalized second order correlation functions are
\begin{equation}
g_{2}^{(\vec{k})}(\tau )=1+e^{-\gamma _{\vec{k}}\tau }[\cos
(E(\vec{k})\tau )+\frac{\gamma _{\vec{k}}}{2E(\vec{k})}\sin
(E(\vec{k})\tau )]^{2} \label{g2n}
\end{equation}
and
\begin{eqnarray}
g_{2}^{(\pm \vec{k} )}(\tau ) &=&1+e^{-\gamma _{\vec{k}}\tau }\{\frac{E^{2}(\vec{k})+%
\frac{\gamma _{\vec{k}}^{2}}{4}}{\bar{n}^{2}V_{d}^{2}(k)}  \nonumber \\
&&+[\cos (E(\vec{k})\tau )+\frac{\gamma _{\vec{k}}}{2E(\vec{k})}\sin (E(\vec{%
k})\tau )]^{2}\} \label{gpm}
\end{eqnarray}

  It is easy to see that the envelope decaying function is given by
  the exciton decay rate $ \gamma_{\vec{k}} $, while the oscillation
  within the envelope function is given by the Bogoliubov
  quasi-particle energy $  E( \vec{k} ) $. Subtracting Eqn.\ref{gpm}
  from Eqn.\ref{g2n} lead to:
\begin{equation}
  g_{2}^{(\pm \vec{k} )}(\tau ) -g_{2}^{(\vec{k})}(\tau )=
  e^{-\gamma _{\vec{k}}\tau }\frac{E^{2}(\vec{k})+%
  \frac{\gamma _{\vec{k}}^{2}}{4}}{\bar{n}^{2}V_{d}^{2}(k)}
\end{equation}
   So we only draw $ g_{2}^{(\pm \vec{k} )}(\tau ) $ in the Fig.18.

\begin{figure}
\includegraphics[width=7cm]{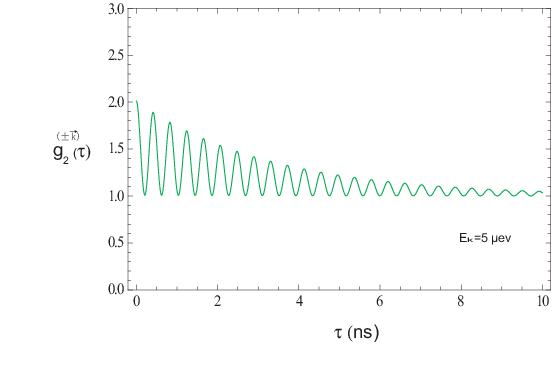}
\caption{ The two photon correlation functions between $ \vec{k} $
and $ -\vec{k} $ against the delay time $ \tau $.  The envelope of
second order correlation function decreases as time evolves
indicates that the emitted photons is photon bunching and the
photo-count statistics is super-Poissonian.}\label{fig15}
\end{figure}

When $\tau =0$ the two photon correlation function are $g_{2}^{(\vec{k}%
)}(0)=2$, so just the mode $ \vec{k} $ alone behaves like a chaotic
light. This is expected because the entanglement is only between $ -
\vec{k} $ and $ \vec{k} $. In fact,
\begin{equation}
g_{2}^{(\pm k )}( 0 )=2+\frac{E^{2}(\vec{k})+\frac{\gamma _{\vec{k}}^{2}}{4}%
}{\bar{n}^{2}V_{d}^{2}(\vec{k})} > g_{2}^{(\vec{k})}(0)=2 .
\end{equation}
  So it violates the classical Cauchy-Schwarz inequality which is
  completely due to the quantum nature of the two mode squeezing
  between $ \vec{k} $ and $ - \vec{k} $.

The normalized  two photon correlation function $ g_{2}^{(\pm
\vec{k} )}(\tau )$ is shown in Fig. 15. From Fig. 15, we can find
that the two photon correlation functions decrease as time interval
$\tau $ increases which suggests quantum nature of the emitted
photons is photon bunching and the photo-count statistics is
super-Poissonian. See Fig.17 for its experimental measurement.

\section{ Discussions on available experimental data and
   possible future experiments}

We will determine how all the important parameters in our theory can
be precisely measured by the experiments. Then we will compare our
results on the EDC with the available experimental data
\cite{butov,snoke} and then discuss possible future experimental
set-ups to detect all the other theoretical predictions achieved in
the previous sections.

\begin{figure}
\includegraphics[width=6cm]{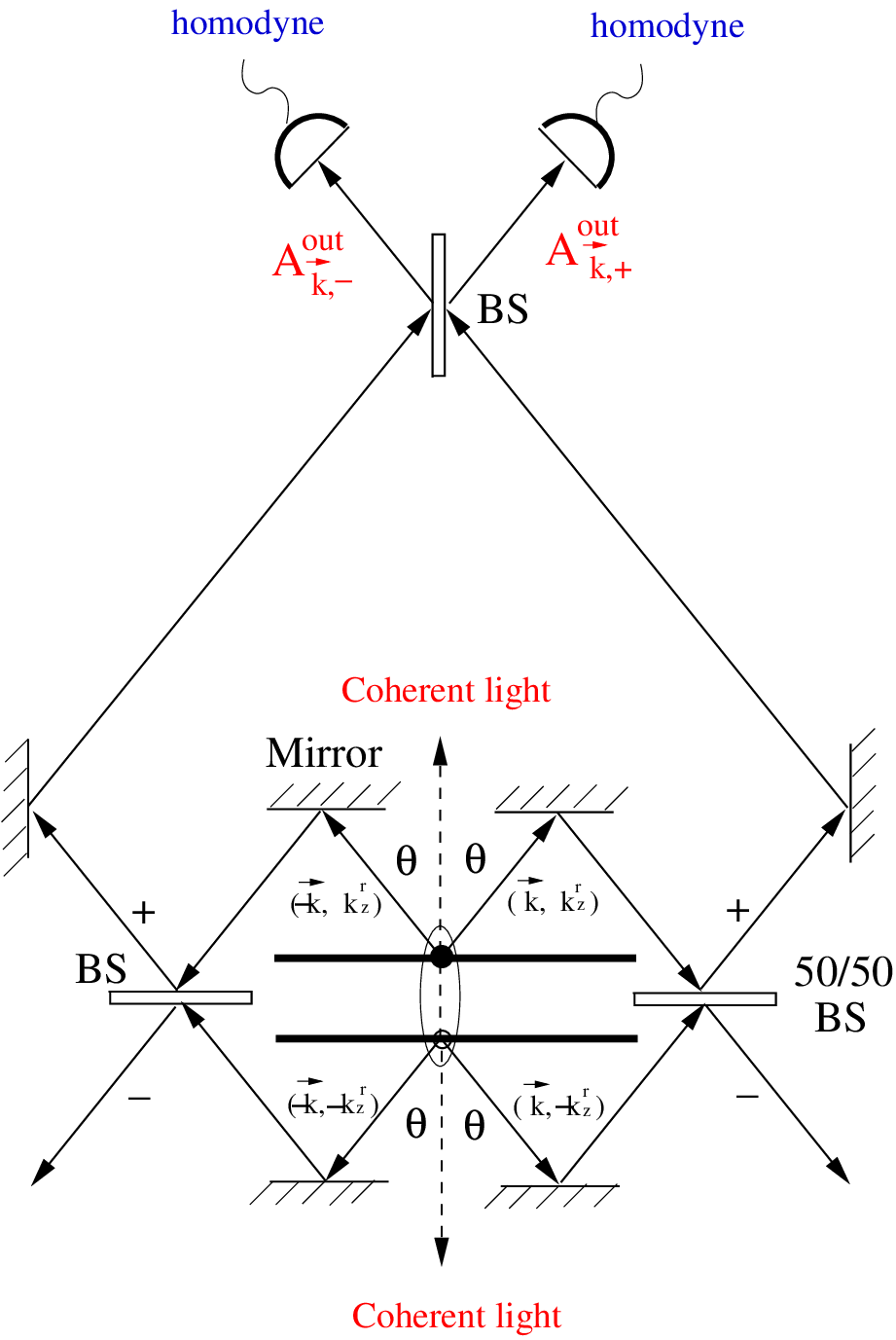}
\caption{ The experimental set-up for the homodyne detection of the
two modes squeezed photon in the EHBL system. The coherent photons
are emitted along the z-direction reflecting the nature of the
condensation. The squeezed photons are emitted along all the tilted
directions reflecting the fluctuation above the condensation. There
is a macroscopic superradiance when the angle $ \theta $ is small.
The homodyne  coming out of $ A^{out}_{\vec{k},+} $ or $
A^{out}_{\vec{k},-} $ is connected to Fig.10 for the single mode
squeezing spectrum measurement. }\label{fig16}
\end{figure}

{\sl 1. Comments on current experimental data }

In \cite{butov,snoke}, the spatially and spectrally resolved
photoluminescence intensity has a sharp peak at the emitted photon
energy $ E=1.545eV $ with a width $ \sim 1 meV  $ at the lowest
temperature $ \sim 1 K \sim 0.1 meV $. The radiation power $ P_{ex}
$ varies from $ 17 \mu W $ to $ 1.6 m W $. The gate voltage is set
at $ V_{g}=1.2 V $. As shown in Eqn.\ref{fixed}, one can identify $
E= \mu= E_{g}-E_{b} + \bar{n} V_{d}(0) =1.545eV $. The energy
conservation at $ k_z=0 $ gives the maximum in-plane momentum $
\vec{k}_{max} \sim \mu /v_{g} \sim 4.3\times 10^{4} cm^{-1} $ where
we used the speed of the light  $ v_{g}\simeq 8.7\times 10^{9}cm/s$
in $ GaAs $ \cite{butov}. Then the maximum exciton energy $ E_{max}=
u k_{max} \sim 0.15meV $ where we used the spin wave velocity $ u
\sim 5 \times 10^{5} cm/s $. Taking the mass of the exciton $ M \sim
0.22m_{0} $ \cite{butov}, then from the expression of the spin wave
velocity  $u =\sqrt{\bar{n}V_{d}(0)/M }$ calculated in
Eqn.\ref{sound}, we find $\bar{n}V_{d}(0)\simeq 0.05meV$ which is
the value we used in all the previous Figs.7-15. Taking the exciton
density $ n \sim 10^{10} cm^{-2} $, we can see that $ K= 2 \pi/a
\sim 10^{6} cm^{-1} $ where $ a $ is the average spacing between
excitons. The average lifetime of the indirect excitons in the EHBL
\cite{butov,snoke} is $ \tau_{ex}\sim 40ns $, then we can estimate
the exciton decay rate $ \gamma_{k} \sim 1/ 40 ns \sim 10^{-4} meV
=0.1 \mu ev $.  At the boundary of the two regimes in the Fig.6
where $E(k^{\ast })= u k^{\ast }= \gamma _{k^{\ast }}/2=10^{-4}
meV$, we can extract $ k^{\ast }=2.4\times 10^{2}cm^{-1}$. So there
are three widely separated momentum scales, $ k^{\ast} \sim 10^{2}
cm^{-1} \ll k_{max} \sim 10^{4} cm^{-1} \ll K \sim 10^{6} cm^{-1}
\sim k_{r} $ which is the roton minimum in Fig.4. So all the
important parameters such as the chemical potential $ \mu $, the
quasi-particle energy $ E( \vec{k} ) = u | \vec{k} | $, the exciton
decay rate $ \gamma_{\vec{k}} $ and the exciton dipole-dipole
interaction strength $ \bar{n}V_{d}(0) $ in our theory can all be
extracted from experimental data.

The typical trap size ( or exciton cloud size )  $ L \sim 30 \mu m
$. The number of excitons is $ N= nL^{2} \sim 10^{5} $ which is
comparable to the number of cold atoms inside a trap in most cold
atom experiments.  The central peak due to the condensate in the MDC
Fig.12 is broadened to $ k_{0} \sim 1/L \sim 10^{3} cm^{-1} $ which
is already larger than the half width due to the quasi-particle $
k^{\ast} \sim 10^{-2} cm^{-1} $. So it is impossible to distinguish
the bi-model structure in the MDC at such a small exciton size. The
coherence length was measured in \cite{coherence}. From
Eqn.\ref{coh}, we find the coherence $ \xi \sim 1/k^{*}\sim 40 \mu m
$ which is slightly larger than the exciton cloud size $ L \sim 30
\mu m $.

For an inhomogeneous condensate $ \langle b ( \vec{k} )  \rangle =
\psi( \vec{k} ) $ inside a harmonic trap $ V(r)=\frac{1}{2} u r^{2}
$, we assume local density approximation (LDA) is valid, then
Eqn.\ref{zero} should be replaced by:
\begin{equation}
i\partial_{t}a( \vec{k}, k_{z} ) =( \omega( \vec{k},k_{z} )-\mu +
V(r) -i\kappa ) a( \vec{k}, k_{z} )-ig^{\ast }(\vec{k}, k_{z}) \psi(
\vec{k} ) \label{inhom}
\end{equation}%
where we have still dropped the zero mode fluctuation of the
inhomogeneous condensate. Because $ g^{\ast }(\vec{k}, k_{z}) \sim
L^{-1/2}_{z} \rightarrow 0 $ and $ \kappa = v_{g}/L_{z} \rightarrow
0 $ such that a stationary state is reached where the energy $
\omega( \vec{k},k_{z}) $ is pinned at the local chemical potential $
\mu(r)= \mu-V(r) $, so the central peak due to the condensate in the
EDC in the Fig.13a is broadened simply due to the change of the
local chemical potential from the center to the edge of the trap $
\Delta \omega= \frac{1}{2} u L^{2} \sim 0.1 meV $ where we used the
experimental value \cite{butov,snoke} $ u \sim  10^{-12} eV nm^{-2}
$. This geometrical broadening due to the trap is already much
larger than the half width due to the quasi-particle $ w_{e} \sim
\gamma_{\vec{k}} \sim 0.1 \mu eV $ in Fig.13b, so it is impossible
to distiguish the bi-modal structure in the EDC in the Fig.13b.
 In order to understand if the observed EDC peak with the width $
\sim 1 meV $ at the lowest temperature $ \sim 1.7 K $  in the
experiments in \cite{butov,snoke} is indeed due to the exciton
condensate, one has to study how the bi-model structures shown in
Fig.13b will change inside a harmonic trap at a finite temperature $
\sim 1.7 K $ and the effects of both dark and bright excitons. This
will be discussed in a separate publication \cite{diffusion}.

{\sl 2. Possible future angle resolved experiments }

As shown  in section VII, the MDC, especially the EDC smear out all
the fine structures of the angle resolved power spectrum in the
Fig.11. So in order to test the theoretical results precisely, it is
necessary to perform the ARPS
measurement. As one rotates the angle $\tan \theta _{k}=\left\vert \vec{k}%
/k_{z}\right\vert $ of the photo-detector in the Fig.1b, one should
see the following interesting behaviors. When $ |\vec{k} | <
|\vec{k}^{\ast } | $, the ARPS and the squeezing spectrum only have
one peak at the resonance frequency $\omega _{k}=\mu \sim 1.545eV$
as shown in Fig.11, Fig.7 and Fig.18b. The line width of the peak is
uniquely determined by the quasi-particle
excitation $u \left\vert \vec{k}\right\vert $ and the decay rate $%
\gamma _{\vec{k}}$.  The one peak will start to split into
the two peaks at $ u \left\vert \vec{k}^{\ast }\right\vert \mathbf{=}\gamma _{\vec{k}%
^{\ast }}/2$ corresponding to the angle $ \sin \theta^{*}=
k^{*}/k_{max} \sim 10^{-2} $, so $ \theta^{*} \sim 10^{-2} $.
 When $ \vec{k} > \vec{k}^{\ast }$ ( namely, $ \theta > \theta^{*} $ ), both the ARPS and
the squeezing spectrum have two peaks at the two resonance frequencies $\omega _{k}=v_{g}(k_{z}^{2}+\vec{k%
}^{2})^{1/2}=1.545eV\pm \lbrack u^{2}\left\vert \vec{k}\right\vert
^{2}-\gamma _{\vec{k}}^{2}/4]^{1/2}$ as shown in Fig.11, Fig.8 and
and Fig.18b. The position and the width of the two peaks are
uniquely determined by the quasi-particle excitation $u k$ and the
decay rate $\gamma _{\vec{k}}$. So all the characters of the
condensate and the fluctuations above it are reflected in the angle
resolved measurements of squeezing spectrum and the power spectrum.

{\sl 3. Possible future two modes phase sensitive homodyne
experiments }

  As shown in the appendix C, due to the relation Eqn.\ref{conn}, the
  experimental set-up in Fig.16 combined with the single mode phase sensitive homedyne set-up in Fig.10
   can measure the squeezing spectrum
  in Eqn.\ref{squ} and the squeezing angle in Eqn.\ref{angle}.

{\sl 4. Possible future two modes HanburyBrown-Twiss type of
experiments }

  The single mode two photon correlation functions  $ g^{(\vec{k})}_{2}(\tau) $ in Eqn.\ref{g2} can be measured
  by the usual HBT set-up \cite{book1,book2}. It is not very
  interesting anyway. From the relation in Eqn.\ref{kz},
  we can see the experimental set-up in Fig.17 can measure the most interesting two modes two photon
  correlation functions $ g^{(\pm \vec{k})}_{2}(\tau) $ in Eqn.\ref{gpm} and Fig.15.

\begin{figure}
\includegraphics[width=5cm]{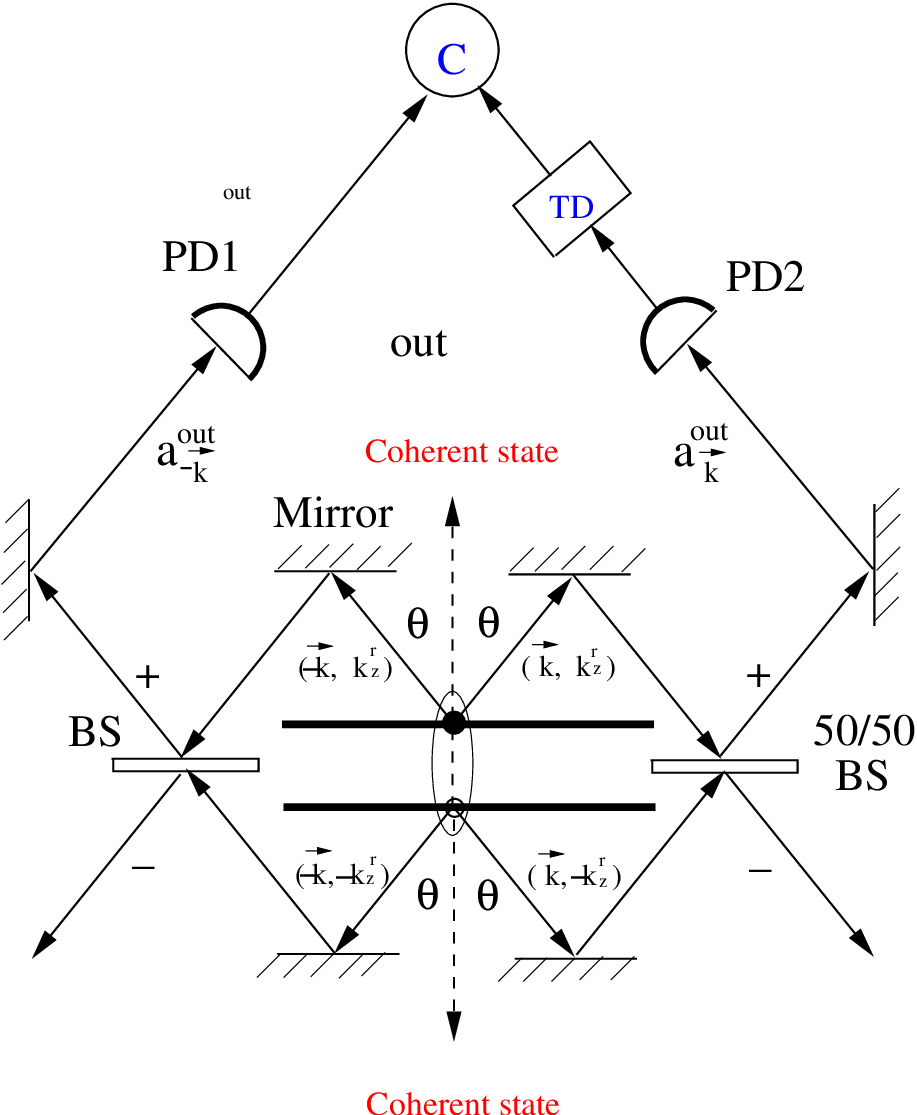}
\caption{  HanburyBrown-Twiss ( HBT ) type of experiment to measure
the two modes two photon correlation functions $ g^{(\pm
\vec{k})}_{2}(\tau) $ in Eqn.\ref{gpm} and Fig.15. } \label{fig17}
\end{figure}

\section{Conclusions}

In conventional quantum optics, coherent light was produced by
stimulated radiations from a pump which leads to particle number
inversion and amplified by optical resonator, here in EHBL, the
coherent state along the normal direction is due to a complete
different and new mechanism: spontaneous symmetry breaking. It has
the following remarkable properties: (1) highly directional: along
the normal direction (1) highly monochromatic: pinned at a single
  energy given by the chemical potential $ \mu $ (3) high power:
  proportional to the total number of excitons.
  These remarkable properties could be useful to build highly
  powerful opto-electronic device.

In conventional non-linear quantum optics, the generation of
squeezed lights
also requires an action of a strong classical pump and a large non-linear susceptibility $%
\chi ^{(2)}$. The first observation of squeezed lights was achieved
in  non-degenerate four-wave mixing in atomic sodium in 1985
\cite{fourwave}. Here in EHBL, the generation of the two mode
squeezed photon is due to a complete different and new mechanism:
the anomalous Green function of Bogoliubov quasiparticle which is
non-zero only in the excitonic superfluid state. The results
achieved in this paper are robust against any microscopic details.
The applications of the squeezed state include (1) the very high
precision measurement by using the quadrature with reduced quantum
fluctuations such as the $ X $ quadrature in the Fig.2 and 3 where
the squeeze  factor reaches very close to $ 0 $ at the resonances
(2) the non-local quantum entanglement between the two twin photons
at $ \vec{k} $ and $ - \vec{k} $ can be useful for many quantum
information processes.
 (3) detection of possible gravitational waves \cite{grav}.

In conventional quantum optics, $ N $ non-interacting atoms confined
in a small volume interact with photon modes, a super-radiance was
proposed by Dicke. Very recently, there are preliminary experimental
evidence that excitons in assembly of quantum dots may emit Dicke's
superradiance \cite{dots}. In conventional laser, due to random
motions of atoms confined in a volume larger than the wavelength of
the laser beam, the laser power is only proportional to $ N $ when
the pump is above the threshold. So far, no super-radiant laser  has
been achieved. Here the ESF phase is a macroscopic quantum phenomena
in a macroscopic sample ( namely in the thermodynamic limit ), so
the super-radiance emitted from this system has completely different
mechanism than the Dicke model.

In conventional quantum phase transitions, all the quantum phases
and phase transitions are stable phases at equilibrium. For example,
all the possible interesting quantum phases in BLQH mentioned in the
introduction are stable equilibrium phases. However, the quantum
phases in EHBL in Fig.3 are just meat-stable phases, they will
eventually decay through emitting photons, photons are very natural
internal probe of the quantum phases and quantum phase transitions.
The characteristics of the photons such as power spectrum, squeezing
spectrum and photon correlations are completely determined by the
nature of the quantum phases such as the ground state and elementary
excitations. The excitation spectrum in a non-equilibrium superfluid
shown in Fig.6b is also different from that in a conventional
equilibrium superfluid shown in Fig.6a. This difference completely
and precisely explained the recent experimental observation of
excitation spectrum of exciton-polariton in a planar microcavity
\cite{expp}.

   In conventional condensed matter experiments, ground states are
completely stable, so in order to  externally  probe the
quasi-particle excitations of a quantum phases, the temperature to
perform the experiments has to be sufficiently high so there are
enough quasi-particle excitations excited above the ground state.
However, the quantum phases in the Fig.3 are just meta-stable, so
the internal probe of emitted photons from the quantum phase can
also reflect the energy spectrum of the quasi-particle even at $T=0
$. This is the salient feature of the internal probe in meta-stable
systems different from the external probes in conventional stable
condensed matter experiments. Exploring the connection between the
quantum phase transitions and quantum optics in meta-stable
non-equilibrium systems is still a developing field and very
exciting and rich.

    It is very instructive to compare the possible exciton BEC in EHBL
    with the well established BEC of ultra-cold atoms \cite{bloch}.
    The quantum degeneracy temperature of two dimensional excitons can be estimated to be $ T^{ex}_{d} \sim 3 K
    $ for exciton density $ n \sim  10^{10} cm^{-2} $ and
    effective exciton mass $ m=0.22 m_{e} $, so it can be reached easily by  $ He $ refrigerator.
    While due to the heavy mass of atoms and the dilute density, $ T^{atom}_{d} \sim  \mu
    K $. Both the exciton BEC and the atomic BEC belong to the weakly interacting
    BEC class, so Bogoliubov theory apply to both cases.
    The BEC to BCS crossover and quantum phases in Fig.3 tuned by
    $ r_{s} $ or imbalance is also quite similar to those of two species of ultra-cold neutral fermions
    tuned by Feshbach resonance or imbalance \cite{bloch}.
    In fact, atomic BEC is also a meta-stable  ground state, it will
    eventually evaporate away. The conventional way to detect the atomic BEC is through
    the time of flight experiments which will destruct the BEC.
     The smoking gun experiment to prove the realization of BEC is through the observation of vortices when the BEC is
  under rotation. In this paper, we
  explicitly showed that although it is hard to rotate the metastable excitonic BEC to generate
  the superfluid vortices,
  the non-equilibrium stationary Bogoliubov quasi-particle spectrum of the exciton BEC
  in teh Fig.6b  can be directly extracted from various
  characteristics of the emitted photons even at $ T=0 $.
  A superradiant behavior was found in the off-resonant light
  scattering  experiment from the BEC condensate \cite{offscattering} and
  studied theoretically in \cite{prltheory}. In the off-resonant
  scattering, the excited atomic state is adiabatically eliminated,
  so important phase information containing the excitation spectrum is lost in
  the  adiabatic elimination procedure. In the exciton BEC phase studied in this
  paper, the photons are always internally resonant with the
  excitons, so it becomes a complete natural internal probe of the ground
  state and excitations of  all the possible  exciton quantum phases.
  However, so far, there is very little experimental
  ways to measure the Bogoliubov quasi-particle of atomic BEC \cite{phononbec}.
  So detecting quantum phases in ultracold atoms remains an outstanding problem.
    Recently, one of the authors and his collaborators developed a theory to detect the nature of quantum phases
    of ultra-cold atoms loaded on optical lattices  by using cavity
    enhanced off-resonant light scattering \cite{jiangming}.
    Excitons carry a electric dipole moment. While in the context of
    cold atoms, very exciting perspectives have been opened by recent
     experiments on cooling and trapping of  $ ^{52}Cr $ \cite{cromium} and
     polar molecules \cite{junpolar}.
     Being electrically or magnetically polarized, the $ ^{52}Cr $  atoms or polar molecules
     interact with each other via long-rang anisotropic
     dipole-dipole interactions. The superfluid and solid phases of
     these polar molecules are under extensive experimental search
     \cite{junpolar,polarroton}.  The excitation  spectrum very similar to that in Fig. 3 including the roton minimum has also been
    proposed in trapped pancake dipolar Bose-Einstein condensates
    \cite{polarroton} in atomic experiments. The presence,
    position and depth of the roton minimum are tunable by varying
    the density, confining potential. So the insights and results achieved in
    this paper may also shed some lights on cold atoms and molecules.

  Before getting to the final summary of the results achieved in
  this paper, we briefly mention some previous and very recent work on semi-conductor electron-hole
  bilayers. BCS pairing of excitons and BEC to BCS crossover were
  discussed in \cite{cross1,cross2}. These mean field calculations
  can only describe the quasi-particle part, but contains no
  information on the collective modes. However, on the BEC side in the Fig.3 discussed in this paper,
  it is the collective mode shown in Fig.6b which contributes to the two mode squeezing state.
  Quantum Monte-carlo using
  trial-wavefunctions on the global phase diagram in Fig. 3 except
  the possible excitonic supersolid phase \cite{ye} were studied in
  \cite{eh1,eh2}. The transport properties were studied in
  \cite{inplane}, but we disagree with the claim made in \cite{inplane} that an in-plane magnetic field can induce
  a counterflow supercurrent.
  The photoluminescence from the
  excitons  inside a harmonic trap was studied by using first order Fermi Golden rules in \cite{golden}
  by treating the exciton BEC as a single two level atom. This
  Golden rule treatment was not able to capture any of the
  physical phenomena explored in this paper which treat the  exciton
  BEC as a quantum phase with its symmetry breaking ground state and
  Bogoliubov excitation spectrum. In the appendix D, we will use a
  first and second order Golden rule calculations which treat the exciton BEC as
  a quantum phase with a ground state and infinitely many excited
  states with different number of Bogoliubov  quasi-particles.
  This kind of Golden calculation based on many particle ground and
  excited states can see some signatures of the results  on coherent state and squeezed state
  achieved by non-perturbative
  calculations in the main context. The two phonon
  squeezing generated by lattice an-harmonic effects  and second-order Raman
  scattering was discussed in \cite{phonon}. As shown in section VII,
  the squeezing is closely related to the macroscopic superradiance, so it would be interesting
  to see if one can achieve a macroscopic super-radiance of phonons.

Several kinds of measurements such as angle resolved power spectrum,
two mode squeezing spectrum, one photon and two photon correlations
functions can completely detect the characters of the emitted
photons which, in turn, are very natural internal probes of the
ground state and excitations of exciton of the quantum phases in the
EHBL system. We established the direct relation between the nature
of quantum phases and the nature of the emitted photons from the
quantum phases: the BEC of excitons lead to the coherent state ( or
a spontaneous laser ) of emitted photons. The anomalous Green
function of the Bogoliubov quasi-particles lead to the two mode
squeezing of emitted photons along all the titled directions. The
whole system behave coherently to emit superradiance even in the
thermodynamic limit. In fact, the ESF phase of the excitons play a
similar role as a  two mode squeezing operator which squeezes the
input vacuum state into a  two mode squeezed state in a suitable
rotated frame. We also evaluated the energy distribution curve (EDC)
and momentum distribution curve (MDC) and compared our results  with
the available experimental data on  MDC and EDC. We determined how
all the important parameters such as the chemical potential $ \mu $,
the quasi-particle energy $ E( \vec{k} ) = u | \vec{k} | $, the
exciton decay rate $ \gamma_{\vec{k}} $ and the exciton
dipole-dipole interaction strength $ \bar{n}V_{d}(0) $ in our theory
can all be extracted from experimental data.   The photons in all
directions show bunching and super-Poissonian. Possible future angle
resolved power spectrum experiments,  two modes phase sensitive
homodyne experiment to measure the squeezing spectrum, two modes
HanburyBrown-Twiss experiment to measure the two photon correlation
functions can be used to directly test these predictions.
In fact, if the exciton BEC was indeed achieved in the past experiments \cite%
{butov,snoke,bell} remains unclear.  In future publications, we will
study the effects of traps, spins of the excitons and finite
temperature. The theoretical results achieved and possible future
experimental set-up proposed in this paper should shed considerable
lights to test if the exciton BEC can be observed in the EHBL
without any ambiguities.


\textbf{Acknowledgements }

We are very grateful for Dr. Peng Zhang for helpful discussions.
J.Ye is indebted to B. Halperin for critical reading of the
manuscript and many critical suggestions. J. Ye is grateful for A.
V. Balatsky, L. Butov,  Jason Ho, Guoxinag Huang, Xuedong Hu, Allan
Macdonald, Hui Deng, Zhibing Li, Qian Niu, Zhe-Yu Oh, Lu Sham, D.
Snoke, Marc Ulrich, Hailing Wang, C. L Yang, Wang Yao, Xiaolu Yu,
Fuchun Zhang, Keye Zhang, Weiping Zhang for helpful discussions. J.
Ye's research at KITP-C is supported by the Project of Knowledge
Innovation Program (PKIP) of Chinese Academy of Sciences, at KITP
was supported in part by the NSF under grant No. PHY-0551164.

\appendix

\section{ The physical pictures in terms of radiation zones }

For simplicity, we first explain the physical picture at the zeroth
order where we neglect the decay of excitons into photons in the
Fig.18a.
Note that at zero temperature $T=0 $, there is no quasi-particle $\beta_{%
\vec{k}} $ excitation in Eqn.\ref{beta}, namely, the input state in
the Fig.4. is the ground state $|BEC> $ ! However, what couples to the photons is $%
\tilde{b}_{\vec{k}} $ in Eqn.\ref{cf2} which is a linear combination of $%
\beta_{\vec{k}} $ and $\beta^{\dagger}_{\vec{k}} $. Equivalently, the phase $%
\theta $ fluctuation used in \cite{ye} is a relativistic excitation
which has both positive particle branch and negative energy
anti-particle branch. The vacuum energy of the photon $E^{ph}_{0}$
is sitting below that of the
exciton $E^{ex}_{0}$ by the chemical potential $\mu $, namely, $%
E^{ex}_{0}-E^{ph}_{0}=\mu $. There exists the resonance condition
not only between the photon and the positive energy branch of the
particle: $\hbar \omega _{k}=1.545eV+u \left\vert \vec{k}\right\vert
$, but also between the photon and the negative energy branch of the
anti-particle: $\hbar \omega _{k}=1.545eV-u \left\vert \vec{k}%
\right\vert $ as shown in Fig.18a.

 When the effects of the excitons decaying into photons are taken into account self-consistently,
 the above picture need to be modified especially at small value of $ \vec{k} $.
 When  $ |\vec{k}|<\vec{k}^{\ast } $, the two quasi-particle branches
 in Fig.18a are not well defined anymore,  the power
 spectrum and the squeezing spectrum only have one peak at resonance
 frequency $\omega _{k}=\mu= 1.545eV$. The one peak starts to
  split into two peaks at $ u \left\vert \vec{k}^{\ast }\right\vert \mathbf{=}\gamma _{\vec{k}%
  ^{\ast }}/2$ where the quasi-particle excitations are still not well defined.
  However, when $ |\vec{k}| \gg |\vec{k}^{\ast }| $, the two
  quasi-particle branches in the Fig.18a remain well defined, so
  both the power spectrum and the squeezing spectrum
  have two peaks at the resonance frequencies $\omega _{k}=v_{g}(k_{z}^{2}+\vec{k%
}^{2})^{1/2}=1.545eV\pm \lbrack u^{2}\left\vert \vec{k}\right\vert
^{2}-\gamma _{\vec{k}}^{2}/4]^{1/2}$ as shown in Fig.18b.
 The flat regime in the Fig.18b precisely and completely explained
 the " anomaly " near $ k=0 $ of the excitation spectrum of
 exciton-polariton in a planar microcavity \cite{expp}.

\begin{figure}
\includegraphics[width=7cm]{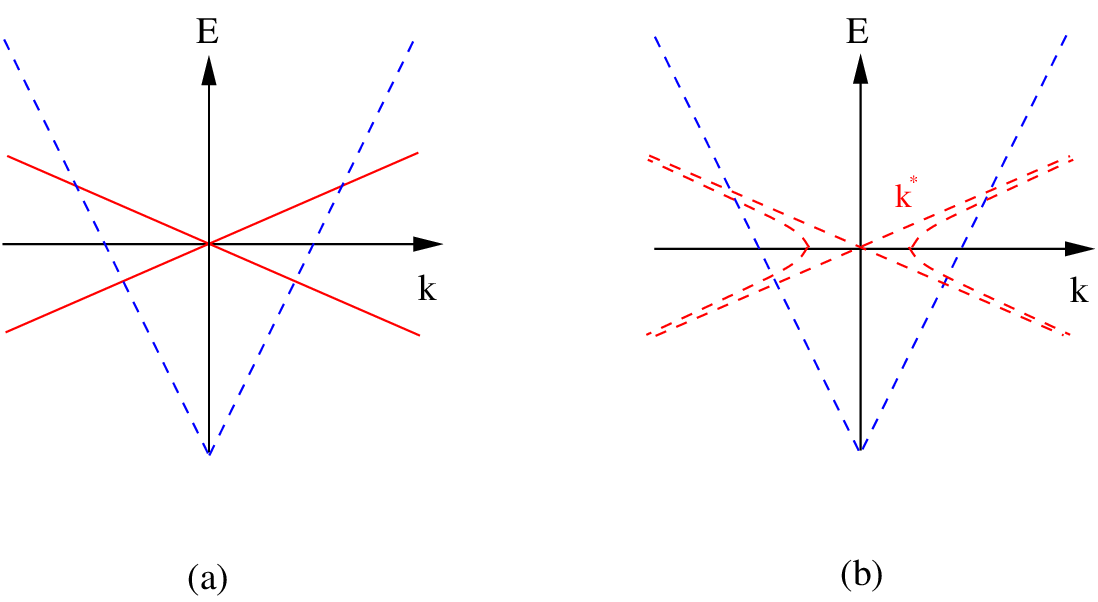}
\caption{ The radiative zone (RZ) in BEC ESF phase (a) The exciton
lifetime $ \tau_{ex} \rightarrow \infty $. The RZ is the regime of
$E(\vec{k})$ between the four intersection points. (b) The exciton
lifetime $ \tau_{ex} $ is large, but finite. The solid gray parabola
lines  denote $ \mu \pm \lbrack u^{2}\left\vert \vec{k}\right\vert
^{2}-\gamma _{\vec{k}}^{2}/4]^{1/2}$ when $ E(\vec{k}) >
\gamma_{\vec{k}}/2 $, the blue dashed line denote $ \omega_{k}=
v_{g} \sqrt{ \left\vert \vec{k}\right\vert ^{2} + k^{2}_{z} } $ at
$k_{z}=0$. When $E (\vec{k})< \gamma _{\vec{k}}/2$, intersection
point between the dashed blue line and the horizontal axis leads to
the maximum squeezing which is a two mode squeezed state, but with
non-zero squeezing angle shown as $ \tilde \pm $ in Fig.5. This
horizontal axis leads to the  the " anomaly " near $ k=0 $ of the
excitation spectrum of exciton-polariton in a planar microcavity
\cite{expp}. When $E (\vec{k})> \gamma _{\vec{k}}/2$, the
intersection points between the dashed blue line and the solid gray
lines lead to the maximum squeezing which is a two mode squeezed
states with zero squeezing angle shown as $ \pm $ in Fig.5. }
\label{fig18}.
\end{figure}

\section{ The two mode squeezed state  from a more intuitive view}

In this appendix, we take a more intuitive way to show that no
matter $E(\vec{k}) <\gamma_{\vec{k}}/2 $ or $E(%
\vec{k}) > \gamma_{\vec{k}}/2 $, the emitted photons are in a  two
mode squeezed state even off the resonance. This conclusion is very
robust and independent of any microscopic details. This appendix
supplement the more formal discussions in section VI.

The position and momentum ( quadrature phase ) operators of the output field
can be more intuitively and straightforwardly defined as:
\begin{eqnarray}
X_{\pm} &=&A_{\vec{k},\pm }^{out}(\omega )+A_{\vec{k},\pm }^{out\dagger
}(-\omega ),  \nonumber \\
Y_{\pm} &=&-i[A_{\vec{k},\pm }^{out}(\omega )-A_{\vec{k},\pm }^{out\dagger
}(-\omega )].
\end{eqnarray}

The Eq. (\ref{bout}) and Eq. (\ref{S}) give the relation $S_{\vec{k}%
,X_{+}}(\omega )=S_{\vec{k},Y_{-} }(\omega )\equiv S_{X}(\omega )$ and $S_{%
\vec{k},Y_{+} }(\omega )=S_{\vec{k}, X_{-}}(\omega )\equiv S_{Y}(\omega )$
and the squeezing spectra:
\begin{eqnarray}
S_{X}(\omega ) &=&1-\frac{2\bar{n}V_{d}(\vec{k})\gamma _{\vec{k}%
}^{2}\epsilon _{\vec{k}}}{\Omega ^{2}(\omega )+\gamma _{k}^{2}E^{2}(k)},
\nonumber \\
S_{Y}(\omega ) &=&1+\frac{2\bar{n}V_{d}(\vec{k})\gamma _{\vec{k}%
}^{2}(\epsilon _{\vec{k}}+2\bar{n}V_{d}(\vec{k}))}{\Omega ^{2}(\omega
)+\gamma _{\vec{k}}^{2}E^{2}(\vec{k})},  \label{off}
\end{eqnarray}%
where $\Omega (\omega )=\omega ^{2}-[E^{2}(\vec{k})-\gamma _{\vec{k}}^{2}/4]$%
.

The two variance functions are defined as \cite{book1}:
\begin{equation}
V_{\pm}(\omega)=< X_{\pm}(\omega) Y_{\pm}(-\omega) + Y_{\pm}(\omega)
X_{\pm}(-\omega) >_{in}/2
\end{equation}

From Eqn.\ref{bout}, we can determine:
\begin{equation}
V_{+}(\omega)= -V_{-}(\omega) = -\frac{2\gamma _{\vec{k}}\bar{n}V_{d}(\vec{k}%
)[\omega ^{2}-E^{2}(\vec{k})+\frac{\gamma _{k}^{2}}{4}]}{\Omega ^{2}(\omega
)+\gamma _{\vec{k}}^{2}E^{2}(\vec{k})}  \label{var}
\end{equation}

From Eqn.\ref{off}, we can find that:
\begin{equation}
S_{X}(\omega) S_{Y}(\omega)= 1+ \frac{ \Omega^{2}(\omega) ( 2 \bar{n} V_{d}(%
\vec{k}) \gamma_{\vec{k}})^{2} }{ [ \Omega^{2}(\omega)+ \gamma^{2}_{\vec{k}}
E^{2}(\vec{k} )]^{2}} \geq 1  \label{offsqu}
\end{equation}
which seems suggest that the state may not be a minimum uncertainty
state. However, from Eqn.\ref{offsqu} and Eqn.\ref{var}, we can see
that $\Delta X^{2}_{\pm} \Delta Y^{2}_{\pm}= 1 + | V_{\pm} |^{2} $
even off the resonance which indicates it is a two mode squeezed
state if we can choose quadrature phase properly. From Eqn.\ref{off}
and Eqn.\ref{var}, we can see that if we define: $\tilde{X}_{\pm}+
i\tilde{Y}_{\pm} = ( X_{\pm} +i Y_{\pm}) e^{\pm i \phi} $ and choose
squeezing parameter $\epsilon= r e^{i 2 \phi} $ such that:
\begin{eqnarray}
\cosh 2r & = & 1+ \frac{ 2 ( \bar{n} V_{d}(\vec{k}) \gamma_{\vec{k}})^{2} }{
\Omega^{2}(\omega)+ \gamma^{2}_{\vec{k}} E^{2}(\vec{k} ) },  \nonumber \\
\cos 2\phi & = & \frac{ \gamma_{\vec{k}}( \epsilon_{\vec{k}} + \bar{n} V_{d}(%
\vec{k}) ) } {\sqrt{ [ \Omega^{2}(\omega)+ \gamma^{2}_{\vec{k}} E^{2}(\vec{k}%
) ] + ( \bar{n} V_{d}(\vec{k}) \gamma_{\vec{k}})^{2}} }  \label{offrphi}
\end{eqnarray}
Then we can show that $\tilde{S}_{X}= \Delta \tilde{X}^{2}_{\pm}=e^{-2r},
\tilde{S}_{Y}= \Delta \tilde{Y}^{2}_{\pm}=e^{2r} $ and
\begin{equation}
\Delta \tilde{X}_{\pm}\Delta \tilde{Y}_{\pm}=1
\end{equation}
which show that the emitted photons are still in the two mode
squeezed state in the basis of $(\tilde{X}_{\pm},\tilde{Y}_{\pm}) $
even off the resonance after making the rotation $\phi $ in the
original basis $( X_{\pm}, Y_{\pm}) $ as
shown in Fig.6. It is easy to see Eqn.\ref{offrphi} is identical to Eqn.\ref%
{angle} and Eqn.\ref{squ}.

\section{ The relation between the calculated quantities and
Experiment measurable quantities }

 Note that the emitted photons at time $t_{1}$ are described by the
operator $a_{k}(t_{1})$ in Eqn.\ref{io}, so it is necessary to find
the relation between $a_{k}(t_{1})$ and $a^{out}_{ \vec{k}}(\omega
)$ through the Eq. (\ref{io}). The Fourier transformation of
Eq.\ref{io} leads to:
\begin{eqnarray}
a_{\vec{k}}^{out}(\omega ) =-\sum_{k_{z}} \frac{1}{
\sqrt{D_{\vec{k}}(\omega+ \mu )}} a_{k}(t_{1})e^{i\omega
t_{1}}\delta (\omega _{k}-\mu -\omega ) \nonumber \\
=-\sqrt{D_{\vec{k}}(\omega  + \mu )}e^{i\omega t_{1}}[a_{(\vec{k}%
,k_{z}^{r})}(t_{1})+a_{(\vec{k},-k_{z}^{r})}(t_{1})]  \label{kz}
\end{eqnarray}
where $k_{z}^{r}=[(\omega +\mu )^{2}/v^{2}_{g} -\left\vert
\vec{k}\right\vert^{2}]^{1/2}$.  So $a_{\vec{k}%
}^{out}(\omega )$ is a linear combination of
$a_{(\vec{k},k_{z}^{r})}(t_{1})$ and
$a_{(\vec{k},-k_{z}^{r})}(t_{1})$, both are at the output time $
t_{1} $.

 Then we can get:
\begin{eqnarray}
 \langle a^{\dagger}_{(\vec{k},k_{z} )}(t_{1}) a_{(\vec{k},k_{z}^{\prime})}(t_{1})
 \rangle  =  \langle a_{\vec{k}}^{\dagger out}(\omega )a_{\vec{k}}^{out}(\omega^{\prime}
 ) \rangle /D_{\vec{k}}(\mu )  \nonumber \\
  =  S_{1}( \vec{k}, \omega ) \delta( \omega
 -\omega^{\prime} )/D_{\vec{k}}(\mu )
\end{eqnarray}
  where, in fact, only $ k_{z}^{\prime}= k_{z} $ contributes and $S_{1}( \vec{k}, \omega
  ) $ is given by Eqn.\ref{power}.
  The total photon number at a given in-plane momentum $ \vec{k} $ is :
\begin{eqnarray}
 \sum_{k_z} \sum_{ k^{\prime}_{z} }
 \langle a^{\dagger}_{(\vec{k},k_{z} )}(t_{1}) a_{(\vec{k},k_{z}^{\prime})}(t_{1})
 \rangle      \nonumber  \\
 = \int d \omega_k D_{\vec{k}}( \omega_k )S_{1}( \vec{k}, \omega )
 = \sum_{k_z} S_{1}( \vec{k}, \omega )
\end{eqnarray}
  which is nothing but the momentum distribution curve in
  Eqn.\ref{mdc}.

 The total photon number at a given energy $ \omega= v_{g} \sqrt{ \vec{k}^{2} + k^{2}_{z} } $ is :
\begin{eqnarray}
 \sum_{\vec{k}} \sum_{k_z} \sum_{ k^{\prime}_{z} }  \delta_{ k_{z},k^{\prime}_{z} }
 \langle a^{\dagger}_{(\vec{k},k_{z} )}(t_{1}) a_{(\vec{k},k_{z}^{\prime})}(t_{1})
 \rangle      \nonumber  \\
 =  \sum _{\vec{k}} \sum_{k_z} S_{1}( \vec{k}, \omega ) \delta( \omega
 -\omega^{\prime} )/D_{\vec{k}}(\mu )
 = \sum _{\vec{k}} S_{1}( \vec{k}, \omega )
\end{eqnarray}
  which is nothing but the energy distribution curve in
  Eqn.\ref{edc}.

 A putative single mode squeezing spectrum of
$a_{\vec{k}}^{out}(\omega )$  can be achieved
 by coincident detection of two photons both  with in-plane momentum
$\vec{k}$, but one with $z$-direction momenta $ k_{z} $, the other
with $ -k_{z}$ shown in Fig.16. As shown in sect V-A, in fact, the
squeezed state is a two mode squeezed state between $ \vec{k} $ and
$ -\vec{k} $, so only the photons described by
$A_{\vec{k},\pm}^{out}(\omega )=[a_{\vec{k}}^{out}(\omega )\pm
a_{-\vec{k}}^{out}(\omega )]/\sqrt{2} = -\sqrt{D_{\vec{k}}(\omega  +
\mu )/2 }e^{i\omega t_{1}} \tilde{A}_{\vec{k},\pm}^{out}( \omega ) $
are in the squeezed state where
\begin{eqnarray}
\tilde{A}_{\vec{k},\pm}^{out}( \omega ) & = &  [
a_{(\vec{k},k_{z}^{r})}(t_{1})+a_{(\vec{k}%
,-k_{z}^{r})}(t_{1})       \nonumber  \\
 & \pm & ( a_{(-\vec{k},k_{z}^{r})}(t_{1})+a_{(-\vec{k}%
,-k_{z}^{r})}(t_{1}) ) ] \label{conn}
\end{eqnarray}%

So the coincident detection experiment shown in Fig. 15 between the
photons with momentum $(\vec{k},\pm k_{z}^{r})$ and $(-\vec{k},\pm
k_{z}^{r})$  at the output time $ t_{1} $ can lead to the squeezing
spectrum $S_{X}(\vec{k}, \omega )$ in Eqn.\ref{squ}. Just as the
Markov approximation made in the section V, because the squeezing
spectrum in Fig.7 and Fig.8 and power spectrum in Fig.11 are
narrowly peaked at $ \omega =0 $ and $ D_{\vec{k}}(\omega + \mu ) $
is also a slowly varying function of $ \omega $, so we can simply
set  $ D_{\vec{k}}(\omega  + \mu ) \sim D_{\vec{k}}( \mu ) $ as a
constant prefactor at a given in-plane momentum $ \vec{k} $ in the
squeezing,  power spectrum, macroscopic super-radiance and
correlation function calculations in the section V. The photons
described by $A_{\vec{k},-}^{out}(\omega )$ is squeezed along the
direction normal to the squeezed direction of the photons described by $A_{\vec{k}%
,+}^{out}(\omega )$ in the phase space as shown in Fig.5.

\section{ Golden rule calculation to second order and two mode squeezed photons }

 This appendix was inspired by J.Ye's visit to Univ.
of Texas at Austin. J.Ye acknowledge very inspiring discussions with
A. Macdonald, Qian Niu and Yao Wang. There are also independent
unpublished work by Jung-Jung Su and A. Macdonald \cite{private}.
  The input-output formalism used in the main text is to solve a integral-differential
  equation under the standard Markov approximation
  which is assuming the coupling constant varies little over the
characteristic energy which is the exciton energy $ \mu \sim 1.54 ev
$, this approximation is valid here.  So the coupling constant $ g_k
$ was treated non-perturbatively under the Markov approximation in
the main text. To reproduce the results achieved in this formalism
by Golden rule is very diffucult, although by pushing Golden rule to
second order, one can see some features on two modes squeezing as
shown in this appendix.

\subsection{ First order Golden rule calculation}

{\sl 1. The condensate at $ \vec{k}=0 $ }

  As shown in the main text, the initial state is taken to be
  $ |i \rangle = |BEC \rangle |0 \rangle $ with energy $
 E_i= 0 $. Due to the in-plane momentum conservation, the final state
 contains one photon is
 $ | f \rangle =  |BEC \rangle |1_{k_z} \rangle $
 with energy $ E_f= \omega_{k_{z}} - \mu  $.
 The transition probability from the initial state to the final
 state can be calculated by the first order Golden rule:
\begin{eqnarray}
  W_{0} & =  & \frac{2 \pi }{\hbar} \sum_{f} | \langle i | H^{\prime} | f
  \rangle |^{2} \delta ( E_f-E_i )    \nonumber  \\
  &  = &  \frac{2 \pi }{\hbar} N \sum_{k_{z}}  g^{2} (k_z)
  \delta( \omega_{k_z} - \mu  )  \nonumber  \\
  & = &  \frac{2 \pi }{\hbar} N \gamma_{0}
\label{gold0}
\end{eqnarray}
   where $ H^{\prime} $ is given by Eqn.\ref{cf2} and
   the exciton decay rate $ \gamma_{0} = |g^{2}(k_z) D $ below Eqn.\ref{npht}.

   From the Eqn.\ref{gold0}, we can get the radiation rate from the condensate:
\begin{equation}
   P^{0}_{rd}= N \gamma_{0} \mu
\end{equation}
   which is the same as Eqn.\ref{power0}.

  It is easy to see that the first order Golden rule calculation is only a one photon process, but the photon
  coherent state  discussed in section IV is a superposition of many photon number states, so it may be
  interesting to push the Golden rule calculation to second order.
  Note that the first order Golden
  calculation in \cite{golden} just treated the exciton BEC as a two level atoms and
  did not use the BEC many body quasi-particle basis.

{\sl 2. The quasi-particles at $ \vec{k} \neq 0 $ }

 As shown in the main text, the initial state is taken to be
  $ |i \rangle = |BEC \rangle |0 \rangle $ with energy $
 E_i= 0 $. Due to the in-plane momentum conservation, the final state
 contains one photon and one quasi-particle
 $ | f \rangle = \gamma^{\dagger}_{-\vec{k}} |BEC \rangle |1_{k} \rangle $
 where $ k=( \vec{k}, k_z) $ with energy $ E_f= \omega_k - \mu + u | \vec{k} | $.
 The transition probability from the initial state to the final
 state can be calculated by the first order Golden rule:

\begin{eqnarray}
  W_{1} & =  & \frac{2 \pi }{\hbar} \sum_{f} | \langle i | H^{\prime} | f
  \rangle |^{2} \delta ( E_f-E_i )    \nonumber  \\
  &  = &  \frac{2 \pi }{\hbar} \sum_{k_{z}}  g^{2} (k)
   | \langle 1_{k} | a^{\dagger}_{k} | 0 \rangle |^{2}
   | \langle BEC | \gamma_{-\vec{k}} \tilde{b}_{\vec{k}} | BEC \rangle |^{2}
                                 \nonumber  \\
   & \times  &  \delta( \omega_k - \mu + u | \vec{k} | )
\label{gold1}
\end{eqnarray}
   where $ H^{\prime} $ is given by Eqn.\ref{cf2}.

   Substituting the inverse of Eqn.\ref{uv}  $
   \tilde{b}_{\vec{k}}= u_{\vec{k}}\gamma_{\vec{k}} -
   v_{\vec{k}}\gamma^{\dagger}_{-\vec{k}} $ into Eqn.\ref{gold1}
   leads to:
\begin{eqnarray}
 W_{1}   & =   &  \frac{2 \pi }{\hbar} \sum_{k_{z}}  g^{2} (k) v^{2}_{\vec{k}}
   \delta( \omega_k - \mu + u | \vec{k} |)  \nonumber  \\
   & = &  \frac{2 \pi }{\hbar} \int d \omega_{k} D_{\vec{k}}( \omega_{k} )  g^{2} (k) v^{2}_{\vec{k}}
   \delta( \omega_k - \mu + u | \vec{k} | )
    \label{gold2}
\end{eqnarray}
    where the photon density of states $ D_{\vec{k}}( \omega_{k} ) $ at a given in-plane momentum
    $ \vec{k} $ is given in Eqn.\ref{dos}. Because $  u |\vec{k} | \ll
    \mu $, then Eqn.\ref{gold2} becomes:
\begin{equation}
 W_{1}    =     \frac{2 \pi }{\hbar}  v^{2}_{\vec{k}} \gamma_{\vec{k}}
\end{equation}
    where the exciton decay rate $ \gamma_{\vec{k}} $ is given in Eqn.\ref{decay}.
    It is easy to see that the photon emission at $ \vec{k} \neq 0 $ is indeed due to the
    quantum depletion in Eqn.\ref{dep}.

    From the Eqn.\ref{gold2}, we can get the radiation rate from the
    quasi-particles at  a given in-plane momentum:
condensate:
\begin{equation}
   P^{rd}_{1}( \vec{k} )=  \frac{2 \pi }{\hbar}  v^{2}_{\vec{k}} \gamma_{\vec{k}} \mu
\end{equation}
   which cab be shown to be the same as
   Eqn.\ref{qradiation} to the linear order in  $ \gamma_{\vec{k}} $.

  It is easy to see that the first order Golden rule calculation is only a one photon process, but the photon
  two mode squeezing
  discussed in section V is a two photon entanglement between $ \vec{k} $ and $- \vec{k} $, so it may be
  interesting to push the Golden rule calculation to second order.  Note that the first order Golden
  calculation in \cite{golden} just treated the exciton BEC as a two level atoms and
  did not use the BEC many body quasi-particle basis.

\subsection{ Second order Golden rule calculation }

The initial state is still taken to be $ |i \rangle = |BEC \rangle
|0 \rangle $ with energy $
 E_i= 0 $. Due to the in-plane momentum conservation, the intermediate state
 $ | m \rangle $ contains one photon and one quasi-particle
 $ | m \rangle = \gamma^{\dagger}_{-\vec{k}} |BEC \rangle |1_{k} \rangle $
 with energy $ E_m= \omega_k - \mu + u | \vec{k} | $.
 AS shown in the last subsection, the matrix element between the initial $ |i \rangle $
 and the intermediate state $ | m \rangle $
 is $ V_{mi}= \langle m | H^{\prime} | i \rangle = - i g(k)
 v_{\vec{k}} $. There are two possible final states (1)
 two photons, but no quasi-particles: $ | f1 \rangle= |BEC \rangle |1_{k } \rangle |1_{-k }
 \rangle $ where  $ k= ( \vec{k}, k_{1z}), -k = ( -\vec{k}, k_{2z}) $
 with energy $ E_{f}= \omega_k - \mu + \omega_{-k} - \mu $.
 The matrix element between the intermediate  and the final state
 is $ V_{fm}= \langle f1 | H^{\prime} | m \rangle = i g(-k)
 u_{\vec{k}} $
 (2) two photons and two quasi-particles:
 $ | f2 \rangle =\gamma^{\dagger}_{-\vec{q}}  \gamma^{\dagger}_{-\vec{k}} |BEC \rangle |1_{k } \rangle |1_{q }
 \rangle $ where  $ k= ( \vec{k}, k_{z}), q = ( \vec{q}, q_{z}) $
 with energy $  E_f= \omega_k - \mu + u | \vec{k} |  +  \omega_q - \mu + u | \vec{q} |
 $. The matrix element between the intermediate  and the final state
 is $ V_{fm}= \langle f2 | H^{\prime} | m \rangle = -i g( q )
 v_{\vec{q}} $.

 The transition probability from the initial state to the final
 state can be calculated by the second order Golden rule calculation.
 The coefficient in front of the final state $ | f \rangle $  is \cite{japan}:
\begin{eqnarray}
  C^{(1)}_{f} & = & 0    \nonumber  \\
  C^{(2)}_{f}   & =  & \frac{ i }{\hbar} \sum_{m} \frac{V_{fm} V_{mi}}{
  E_m - E_i } \int^{t}_{0} [ e^{i \omega_{fi} t^{\prime} }
  - e^{i \omega_{fm} t^{\prime} }] d t^{\prime}
\end{eqnarray}

   When $ E_{m} $ differes from $ E_n $ and $ E_i $, the
   contribution from second term leads to a rapid oscillation which
   does not lead to a transition probability growing with the time $
   t $, so the second term can be dropped. The tarnsition
   probability from the initial state $ i $ to the final state $ f $
   is:
\begin{equation}
  W_{i\rightarrow f} =   \frac{ 2 \pi }{\hbar}| V_{fi}+  \sum_{m} \frac{V_{fm} V_{mi}}{
  E_m - E_i } |^{2}  \delta( E_f-E_i)
\end{equation}
  In the following, we will use this formula to calculate the transition to the state
  $ |f1 \rangle $ and $ |f2 \rangle $ respectively.

\subsubsection{ The condensate at $ \vec{k}=0 $ }

{\sl 1. The transition probability to the state $ |f1 \rangle_0 =
 |BEC \rangle |1_{k_z } \rangle |1_{q_z }\rangle, k_z \neq q_z $ }

  The transition from the initial state $ |i \rangle $ to the final
  state  $ |f1 \rangle_0 $ can be through two possible intermediate
  states  $ | m1 \rangle =  |BEC \rangle |1_{k_z} \rangle $
  and  $ | m2 \rangle =  |BEC \rangle |1_{q_z} \rangle
  $, so the transition probability to the final state $ |f1 \rangle_0 $ after summing over the two
  intermediate states is:
\begin{eqnarray}
 W_{(f1)_0} &  = &    \frac{2 \pi }{\hbar}  \sum_{k_{z} \neq q_z} \sum_{q_{z} }
  N^{2} |  g(k_z ) g( q_z ) |^{2}  | \frac{1}{ \omega_{k_z} - \mu   }
    + \frac{1}{ \omega_{q_z} - \mu   } |^{2}
       \nonumber  \\
 & \times &   \delta( \omega_{k_z } - \mu +   \omega_{q_z } - \mu  ) =0
\label{con1}
\end{eqnarray}
   where  the first and second sums are over all the possible final states with $ k_z \neq q_z $,
   the energy conservation
   $ E_f=E_i $, namely, $\omega_{k_z } - \mu +   \omega_{q_z } -
   \mu=0  $ is enforced by the $\delta $ function.

{\sl 2. The transition probability to the state $ |f2 \rangle_0 =
 |BEC \rangle |2_{k_z } \rangle  $ }

  The transition from the initial state $ |i \rangle $ to the final
  state  $ |f2 \rangle_0 $ can be through only one intermediate
  states  $ | m \rangle =  |BEC \rangle |1_{k_z} \rangle $,
  so the transition probability to the final state $ |f2 \rangle_0 $
  is:
 \begin{equation}
 W_{(f2)_0}   =     \frac{2 \pi }{\hbar}  \sum_{k_{z} }
  N^{2} |  g^{2}(k_z ) |^{2}  | \frac{1}{ \omega_{k_z} - \mu   } |^{2}
   \delta( 2( \omega_{k_z } - \mu )  )
\label{con2}
\end{equation}
   where  the  sums is over all the possible final states,
   the energy conservation
   $ E_f=E_i $, namely, $ 2 (\omega_{k_z } - \mu )=0   $ is enforced by the $\delta $ function.

   From Eqns.\ref{con1} and \ref{con2}, one can see that the emitted
   photons can only have a single momentum pinned at the chemical
   potential $ \mu $ which is consistent with the results achieved in
   the section  IV. In order to treat the pole structure in Eqn.\ref{con2}, one
   has to treat the condensate-photon by Heisenberg equation of motion as done
   in the section IV.

\subsubsection{ The quasi-particles at $ \vec{k} \neq 0 $ }

{\sl 1. The transition probability to the state $ |f1 \rangle =
 |BEC \rangle |1_{k } \rangle |1_{-k }\rangle $ }

  The transition from the initial state $ |i \rangle $ to the final
  state  $ |f1 \rangle $ can be through two possible intermediate
  states  $ | m1 \rangle = \gamma^{\dagger}_{-\vec{k}} |BEC \rangle |1_{k} \rangle $
  and  $ | m2 \rangle = \gamma^{\dagger}_{\vec{k}} |BEC \rangle |1_{-k} \rangle
  $, so the transition probability to the final state $ |f1 \rangle $ after summing over the two
  intermediate states is:
  \begin{eqnarray}
 W_{(f1)}  & =   &  \frac{2 \pi }{\hbar}  \sum_{k_{z1}} \sum_{k_{z2}}
  |  g(k ) g(-k ) u_{\vec{k}}
  v_{\vec{k}} |^{2}  | \frac{1}{ \omega_k - \mu + u | \vec{k} |  }
     \nonumber  \\
  & + &
   \frac{1}{ \omega_{-k} - \mu + u | \vec{k} |  } |^{2}
   \delta( \omega_{k } - \mu +   \omega_{ - k } - \mu  )
     \nonumber  \\
  &  =  &  \frac{2 \pi }{\hbar} | u_{\vec{k}}
  v_{\vec{k}} |^{2} \int^{2 \mu}_{0} d \omega_{k} D_{\vec{k}}( \omega_{k}
   ) D_{\vec{k}}( 2 \mu- \omega_{k} )     \nonumber  \\
   & \times & ( g_{\vec{k}}( \omega_k) g_{\vec{k}}( 2 \mu - \omega_k
   ))^{2} ( \frac{ 2 u | \vec{k} | }{ ( \omega_{k}- \mu )^{2} - ( u | \vec{k}
   | )^{2} } )^{2}
\label{1a}
\end{eqnarray}
   where  the first and second sums are over all the possible final states, the energy conservation
   $ E_f=E_i $, namely, $ \omega_{k } - \mu +   \omega_{ - k } - \mu =0
   $, is enforced by the $\delta $ function. .
   We can see that there are two resonances at $  \omega_{k} = \mu \pm u | \vec{k}| $.
   We recovered the results achieved in Sec. VI-1 in the weak decay
   case $ E( \vec{k} ) > \gamma(\vec{k})/2 $ where the quasi-particles are well defined. This is expected
   because in the Golden rule calculations, we treat $
   \gamma(\vec{k}) $ perturbatively, so can only discuss the weak
   decay case, the strong decay case is beyond the scope of the
   Golden rule calculation.
   Note that the prefactor
   $ u_{\vec{k}} v_{\vec{k}}= \frac{ \bar{n} V_{d}(\vec{k}) }{ 2 E(\vec{k}) } $ is completely due
   to the exciton dipole-dipole interaction.

   In order to treat the pole structure near the two resonances, one
   has to treat the exciton-photon system self-consistently as done
   in the section V.

{\sl 2. The transition probability to the state
  $ |f2 \rangle = \gamma^{\dagger}_{-\vec{q}}  \gamma^{\dagger}_{-\vec{k}} |BEC \rangle |1_{k } \rangle |1_{q }
 \rangle $ }

  The transition from the initial state $ |i \rangle $ to the final
  state  $ |f2 \rangle $ can be through two possible intermediate
  states  $ | m1 \rangle = \gamma^{\dagger}_{-\vec{k}} |BEC \rangle |1_{k} \rangle $
  and  $ | m2 \rangle = \gamma^{\dagger}_{-\vec{q}} |BEC \rangle |1_{q} \rangle
  $, so the transition probability to the final state $ |f1 \rangle $ after summing over the two
  intermediate states is:
  \begin{eqnarray}
 W_{(f2)}  & =   &  \frac{2 \pi }{\hbar}  \sum_{k_{z}} \sum_{q_{z}}
  |  g(k ) g( q ) v_{\vec{k}}
  v_{\vec{q}} |^{2}  | \frac{1}{ \omega_k - \mu + u | \vec{k} |  }
     \nonumber  \\
  & + &
   \frac{1}{ \omega_{q} - \mu + u | \vec{q} |  } |^{2}
   \delta( \omega_{k } - \mu +  u | \vec{k} |
           +  \omega_{q} - \mu  + u | \vec{q} | )
     \nonumber  \\
  &  =  &  0
\label{2a}
\end{eqnarray}
   where the energy conservation $ E_f=E_i $, namely, $ \omega_{k } - \mu +  u | \vec{k} |
           +  \omega_{q} - \mu  + u | \vec{q} |=0  $ is enforced by the $ \delta $ function,
           We conclude that there is no transition probability to
           the state $ | f2 \rangle $ in sharp contrast to the state $ | f1 \rangle
           $.

\subsection{ Discussions }

  Combining the calculations in the first and  the second order
  Golden rule calculations in the last two subsections, we can write the complete Hilbert space
  of the exciton-photon system at a given inplane momentum $ \vec{k} $ as:
\begin{eqnarray}
  | \Psi \rangle & = &  c_{0}|BEC \rangle | 0 \rangle
    + c_{1} \gamma^{\dagger}_{-\vec{k}} |BEC \rangle |1_{k} \rangle
      \nonumber \\
   &  +  &  c_{2a} |BEC \rangle |1_{k} \rangle |1_{-k} \rangle
              \nonumber  \\
   & +  &  c_{2b} \gamma^{\dagger}_{-\vec{k}}
              \gamma^{\dagger}_{-\vec{q}} |BEC \rangle |1_{k} \rangle |1_{q} \rangle
              + \cdots
\label{many}
\end{eqnarray}
   where the $ \cdots $ stand for multi-quasi-particles or multi-
   photon states. Eqn.\ref{2a} gives $ c_{2b}=0 $.
    It is the $ c_{2a} $ term which leads to the two mode squeezing
    between $ \vec{k} $ and $ - \vec{k} $.  We can
    compare Eqn.\ref{many} with the well known Wegner-Weiskoff theory of the
    spontaneous radiation of two level atoms where the Hilbert space
    is: $  | \Psi \rangle  =  c_{0} |e, 0 \rangle + c_{1 \vec{k} } |
    g, 1_{\vec{k} } \rangle $. Because the
    atom only has two levels: the excited state $ |e \rangle $ and
    the ground state $ |g \rangle $, so the atom+photon system can only has
    either no photon or one photon state. In this case, the Golden
    rule calculation  to first order can gives the correct decay rate.
    In the three level atom system with two photon cascades:
    $  | \Psi \rangle  =  c_{a} |a, 0 \rangle + c_{b, \vec{k} } |
    b, 1_{\vec{k} } \rangle  + c_{c, \vec{k}, \vec{q} } |
    c, 1_{\vec{k} }, 1_{\vec{q} } \rangle $,  the
    atom only has three levels: $ |a \rangle, |b \rangle, |c \rangle  $,
    so the atom+photon system can only has no photon, one photon and two photon states.
    In this case, the Golden
    rule calculation  to second order is needed to give the correct decay rate.
    It is known there are no entanglements between the two photon
    states.

    While the exciton BEC system has infinite many body states with different number of
    quasi-particle excitations, so the exciton-photon system can
    has infinite many photon states as shown in
    Eqn.\ref{many}. This is the crucial difference between a quantum
    phase and a two levels or three levels system. So the Golden rule calculations are  very limited in this case.
    To first order, Golden
rule is just a one photon process, so it fails to capture the
squeezed photons which are a two photon process. The two mode
squeezed photons are the pairing of two photons between $ \vec{k} $
and $-\vec{k} $ just like pairing of two electrons between $ \vec{k}
$  and $-\vec{k} $. In this appendix, we pushed the Golden rule to
the second order which connects the initial and final state by two
photons, we indeed find the signatures of the two photon squeezing.
The formalism used in the main text is an input-output formalism
which is to all order of the coupling constant, the only
approximation  made is the Markov approximation which was also used
in conventional Weisskopf-Wigner theory of spontaneous emission
\cite{book1}. By looking at the squeezing spectrum Eqn.\ref{squ} and
the power spectrum Eqn.\ref{power}, one can see the coupling
constant $ \gamma_{\vec{k}} $ appear in both numerators and
denominators which take  the two photon squeezing into account
completely. The power spectrum which shows the macroscopic
superradiance is also due to the two photon pairing. Especially
below Eqn.\ref{edc}, one can see the coupling constant $
\gamma_{\vec{k}} $ is in the denominator in the MDC curve in the $
\vec{k} \rightarrow 0 $  limit which is in the macroscopic
super-radiance case, so can not be achieved by the perturbation
Golden rule. One need to sum over infinite number of terms to
reproduce the results achieved from the input-output formalism.


\begin{thebibliography}{99}
\bibitem{foot} In fact, Eqns.\ref{g2},\ref{g2pm} are not
time-ordered, so Wick theorem may not apply. However, direct
calculations in $ \omega $ space by using Eqn.\ref{bout}, then
tranforming back to $ \tau $ space give the same answers as those
from applying Wick theorem.
\end{thebibliography}
\end{document}